\documentclass[nofootinbib,superscriptaddress]{revtex4}

\usepackage{graphicx,color}
\def\ifmath#1{\relax\ifmmode #1\else $#1$\fi}
\def\lsim{\mathrel{\raise.3ex\hbox{$<$\kern-.75em\lower1ex\hbox{$\sim$}}}}
\def\gsim{\mathrel{\raise.3ex\hbox{$>$\kern-.75em\lower1ex\hbox{$\sim$}}}}
\def\hsm{h_{\rm SM}}
\def\mhsm{m_{\hsm}}
\def\ls#1{\ifmath{_{\lower1.5pt\hbox{$\scriptstyle #1$}}}}

\newenvironment{Eqnarray}%
     {\arraycolsep 0.14em\begin{eqnarray}}{\end{eqnarray}}

\newcommand{\beq}{\begin{equation}}
\newcommand{\eeq}{\end{equation}}
\newcommand{\beqa}{\begin{Eqnarray}}
\newcommand{\eeqa}{\end{Eqnarray}}

\begin{document}
\bibliographystyle{revtex}  

\begin{flushright}
FERMILAB-Conf-02/053-T \\
MCTP-02-16 \\
SCIPP-02/03 \\
BNL-69066 \\
DESY 02-050 \\  
hep-ph/0203229  \\
March 2002
\end{flushright}

\title{Executive Summary of the Snowmass 2001 Working Group (P1)\\
``ELECTROWEAK SYMMETRY BREAKING''}

\author{Marcela Carena}
\email[]{carena@fnal.gov}
\affiliation{Theoretical Physics Department, Fermi 
National Accelerator Laboratory, Batavia, Illinois 60510-0500, USA}

\author{David W. Gerdes}
\email[]{gerdes@umich.edu}
\affiliation{Department of Physics, University of Michigan, 
Ann Arbor, Michigan 48109-1120, USA}

\author{Howard E. Haber}
\email[]{haber@scipp.ucsc.edu}
\affiliation{Santa Cruz Institute for Particle Physics, 
University of California, Santa Cruz, California 95064, USA}

\author{Andr\'e S. Turcot} 
\email[]{turcot@fnal.gov}
\affiliation{Brookhaven National Laboratory, Upton, New York 11973-5000, USA}

\author{Peter M. Zerwas}
\email[]{zerwas@desy.de}
\affiliation{Deutsches Elektronen-Synchrotron, D-22603 Hamburg, Germany}       
\vspace{3 mm}
\date{\today}

\begin{abstract}
In this summary report of the 2001 Snowmass Electroweak Symmetry Breaking
Working Group, the main candidates for theories of
electroweak symmetry breaking are surveyed, and the criteria for
distinguishing among the different approaches are discussed.
The potential for observing electroweak symmetry
breaking phenomena at the upgraded Tevatron and the LHC is described.
We emphasize the importance of a high-luminosity $e^+e^-$ linear
collider for precision measurements to clarify the underlying
electroweak symmetry breaking dynamics.  Finally, we note the 
possible roles of the $\mu^+\mu^-$ collider and VLHC for further 
elucidating the physics of electroweak symmetry breaking.
\vspace{8 mm}
\end{abstract}

\maketitle

\section{The Origin of Electroweak Symmetry Breaking}

\subsection{Introduction}

Deciphering the mechanism that breaks the electroweak symmetry
and generates the masses of the known fundamental particles
is one of the central problems of particle 
physics~\cite{Higgs-orig,weinberg}.  This mechanism will
be explored by experiments now underway at the upgraded
proton-antiproton Tevatron collider and
in the near future at the Large Hadron Collider (LHC).
Once evidence for electroweak symmetry breaking (EWSB) dynamics is
obtained, a more complete understanding of the mechanisms involved
will require experimentation at future $e^+e^-$ linear 
colliders now under development.  In certain scenarios, a $\mu^+\mu^-$
collider or the next generation of very large hadron colliders after
LHC (VLHC) can play an important role in establishing the
nature of the mass generation mechanism for the fundamental particles.  

The dynamics of electroweak 
symmetry breaking requires the existence of at least one new particle 
beyond the presently
observed spectrum of the Standard Model.  The energy scale associated
with electroweak symmetry breaking dynamics must be of order 1~TeV or
below in order to preserve the unitarity of
the scattering matrix for electroweak gauge bosons~\cite{unitarity},
a principle guaranteed
by quantum mechanics. The specific details of the mechanism
realized in nature to break the electroweak symmetry have 
far-reaching consequences for possible new physics beyond the Standard Model.

The generation of masses for the $W^\pm$ and $Z$ gauge bosons is
associated with the dynamics of electroweak symmetry breaking.
Goldstone bosons, which are 
massless scalar degrees of freedom, are generated by
the symmetry-breaking mechanism and transformed into the longitudinal spin
components of the $W^\pm$ and $Z$.  At present, the underlying nature
of this dynamics is unknown.  Two broad classes of electroweak
symmetry breaking mechanisms have been pursued theoretically.  In one
class of theories, electroweak symmetry breaking dynamics is
weakly-coupled, while in the second class of theories the dynamics is
strongly-coupled.

In theories of weak electroweak symmetry breaking, the symmetry is
broken by the dynamics of a weakly-coupled sector of self-interacting
elementary scalar fields.  These self-interactions give rise to a
non-vanishing scalar field in the vacuum.  Interactions of the
Standard Model fields with this vacuum field generate the masses of
the gauge bosons, quarks and leptons.  In addition, the physical
particle spectrum also contains massive scalars---the Higgs
bosons~\cite{Higgs-orig,hhg}.
All fields remain weakly interacting at energies up to the grand
unification scale which is close to the Planck scale.  At energies at
and beyond the Planck scale, gravitational interactions become as
important as the strong and electroweak interactions, and must be
incorporated in the theory in a consistent quantum mechanical way.  In
the weakly-coupled approach to electroweak symmetry breaking, the
Standard Model is very likely embedded in a supersymmetric 
theory~\cite{susyreview} in
order to stabilize the large gap between the electroweak and the grand
unification (and Planck)
scales in a natural way~\cite{natural}. These theories predict a spectrum
of Higgs scalars~\cite{susyhiggs}, 
with the lightest Higgs scalar mass below about 
135~GeV~\cite{higgslim} in the model's minimal realization.

Alternatively, strong breaking of electroweak symmetry is accomplished
by new strong interactions near the TeV scale~\cite{weinberg,hill02}.  
In most realizations
of this approach, condensates of fermion-antifermion pairs are
generated in the vacuum. The interactions of the electroweak gauge
bosons with the associated Goldstone modes generate the masses of the
gauge bosons.  These models typically possess no elementary scalar
fields.  In some approaches, composite scalar fields, which may
resemble physical Higgs bosons, exist in the spectrum and are composed
of fermionic constituents.  
These constituents may be new matter
fermions, as in the case of technicolor 
models~\cite{weinberg,technireview,technitwo}, 
or a combination of
new heavy quarks and the heavy Standard Model top and bottom quarks,
as in the case of top-color models~\cite{topcolor,topseesaw}.  
Quark and lepton masses are
generated by introducing either effective Yukawa couplings between the
composite scalar fields and the fermion fields or by extending the
system by adding new additional gauge interactions that mediate the
interactions between the Standard-Model fermions and the new fermions.
These theoretical approaches are quite complicated constructs; the
simplest realizations are generally in conflict with experimental
constraints such as precision electroweak data and flavor changing
neutral current bounds.

A new approach to electroweak symmetry breaking has recently been
under intense investigation, in which extra space dimensions beyond
the usual $3+1$ dimensional spacetime are 
introduced~\cite{n12a,extradim,extrasnow} with
characteristic sizes of order $(\rm TeV)^{-1}$.
In such scenarios, the mechanisms for
electroweak symmetry breaking are inherently
extra-dimensional, and can result in a phenomenology significantly
different from the usual approaches mentioned above.
For example,
the mass of the Higgs boson may be generated through interactions
with Kaluza-Klein states in the bulk of multi-dimensional
space-time.  In some cases, the Higgs couplings
to quarks and leptons may be drastically altered compared with the
predictions of the Standard Model~\cite{drastic}.  Some models
exhibit new scalar fields ({\it e.g}., radions) which mix with the
Higgs bosons and can result in significant 
shifts in the Higgs couplings~\cite{extrasnow,radions,moreradions}.
In all such approaches, new physics must be revealed at the TeV scale or below.
Clearly, in order to understand {\it any} theory of electroweak symmetry
breaking dynamics, it is critical to explore 
and interpret the attendant new TeV-scale physics beyond the Standard Model.

\subsection{Criteria for Distinguishing among Models of EWSB}

\begin{figure}[t!]
\begin{center}
\unitlength1cm
\begin{picture}(15,8.3)
\put(-1.0,-0.9){\includegraphics[height=9.80cm,width=8.85cm]{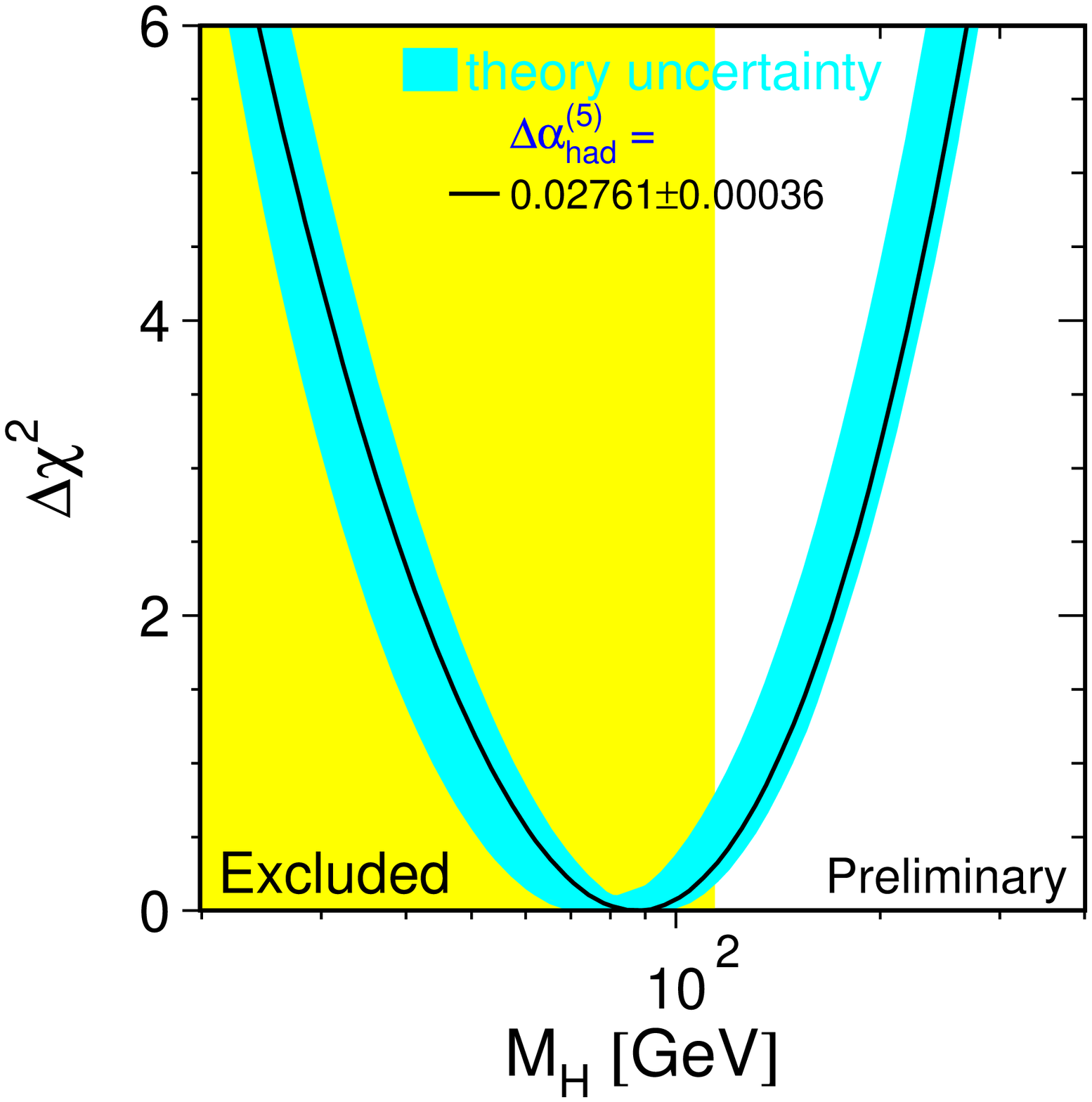}}
\put(8.05,-0.15){\includegraphics[height=8.47cm,width=7.975cm]{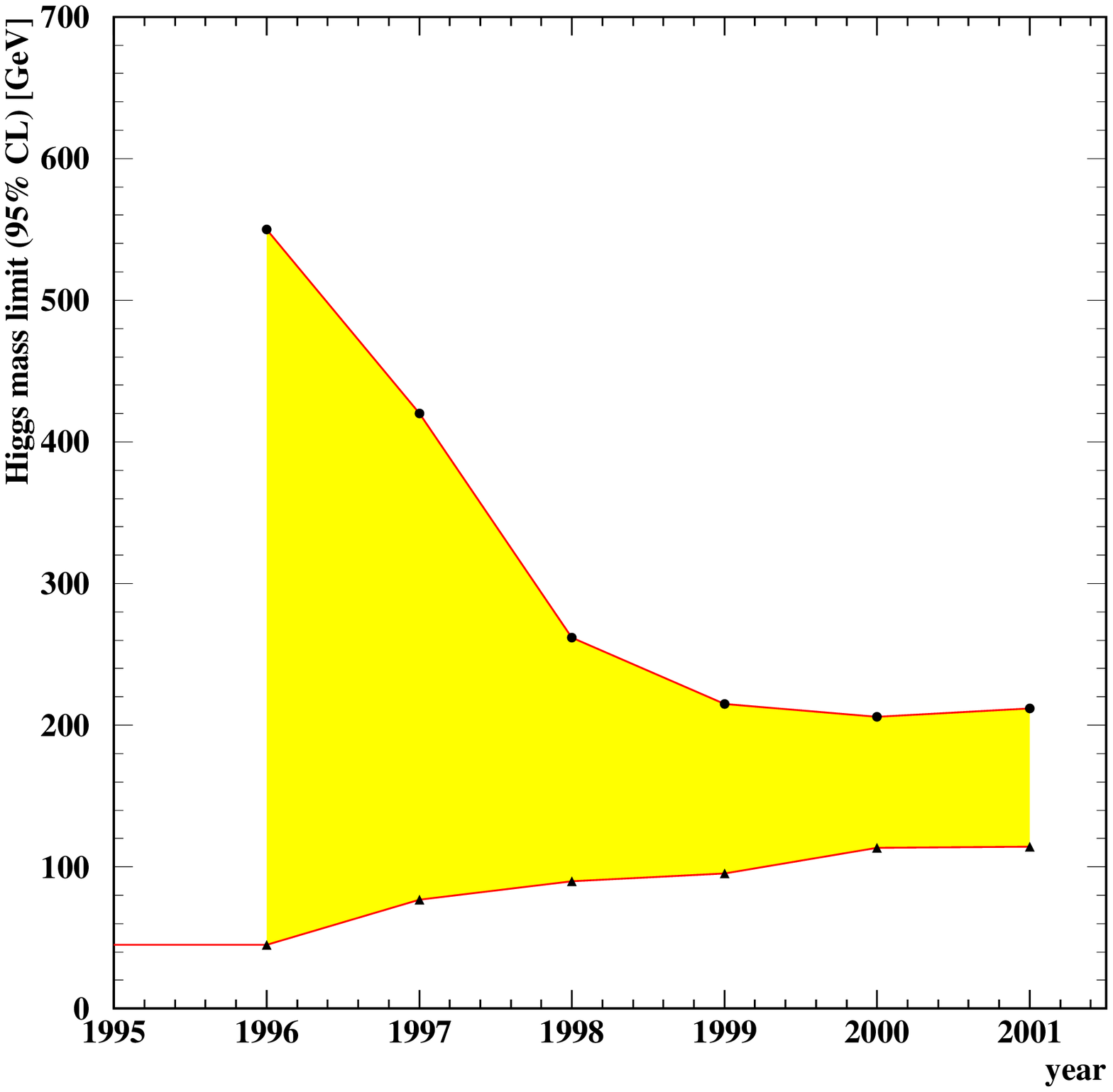}}
\end{picture}
\end{center}
        \caption{(a)~The ``blueband plot'' shows $\Delta
\chi^2\equiv \chi^2-\chi^2_{\min}$ as a function of the Standard Model
Higgs mass~\protect\cite{lepewwg,p1-wg1-rep}.  
The solid line is a result of a global fit 
using all data; the band represents the theoretical
error due to missing higher order corrections.  The rectangular shaded region
shows the 95\% CL exclusion limit on the Higgs mass from direct
searches. (b)~The evolution of
the bounds on the Standard Model Higgs mass from 1996--2001.  The
upper boundary corresponds to the 95\% CL upper bound on the Higgs
mass derived from the global fit to electroweak data, 
and the lower boundary corresponds to the 95\% CL lower bound on the
Higgs mass from direct searches.}
\label{fig:blueband}
\end{figure}

Although there is as yet no direct evidence for the nature of
electroweak symmetry breaking dynamics, present data
can be used to discriminate among the different approaches.  For example, 
precision electroweak data, accumulated in the past decade at LEP, SLC,
the Tevatron, and elsewhere, 
strongly supports the Standard Model with a weakly-coupled 
Higgs boson~\cite{lepewwg,p1-wg1-rep}.
Moreover, the contribution of new physics, which
can enter through $W^\pm$ and $Z$ boson vacuum polarization
corrections, is severely constrained.  This fact has already
served to rule out several models of strongly-coupled electroweak symmetry
breaking dynamics.  The Higgs boson contributes
to the $W^\pm$ and $Z$ boson vacuum polarization through loop effects, and
so a Standard Model fit to the electroweak data yields information
about the Higgs mass. Present fits
indicate that the Higgs mass should be around 100 GeV
[with a fractional $1\sigma$ uncertainty of about $50\%$],
comparable to the direct search upper limit, and must be less than
about 200 GeV at 95\% CL, as shown in Figure~\ref{fig:blueband}a.  
The electroweak data have
improved significantly over the past decade, as shown in 
Figure~\ref{fig:blueband}b, to the
extent that the conclusions of the 2001 Snowmass Workshop
are considerably sharper than what was possible at the end of the 1996 
Snowmass Workshop.

There are some loopholes that can be exploited to circumvent
this conclusion.  It is possible to construct models of new physics
where the goodness of the global Standard
Model fit to precision electroweak data is not compromised
while the strong upper limit on the Higgs mass is
relaxed.  In particular, one can construct effective 
operators~\cite{newoperators}
or specific models~\cite{Peskin-Wells,topseesaw,moreradions}
of new physics where the Higgs mass is significantly heavier, but the new
physics contribution to the
$W^\pm$ and $Z$ vacuum polarizations is still consistent with the
experimental data.  In addition, some have argued that the 
global Standard Model fit exhibits 
possible internal inconsistencies~\cite{chanowitz}, which would suggest that
systematic uncertainties have been underestimated and/or new physics
beyond the Standard Model is required.  
Thus, although weakly-coupled electroweak
symmetry breaking models seem to be favored
by the strong upper limit on the Higgs mass, one cannot definitively
rule out all other approaches.  

Nevertheless, one additional piece of data is very suggestive.  Within
the supersymmetric extension of the Standard Model, grand unification
of the electromagnetic, the weak and the strong gauge interactions can
be achieved in a consistent way, strongly supported by the prediction
of the electroweak mixing angle at low energy scales with an accuracy at the
percent level~\cite{IbanezRoss,susygut}.
The significance of this prediction is not easily matched by other
approaches.  For example, in strongly-coupled electroweak symmetry breaking
models, unification of couplings is not addressed {\it per se}, whereas in  
extra-dimensional models it is often achieved by introducing new 
structures at intermediate energy scales.
Unless one is willing to regard
the apparent gauge coupling unification as a coincidence, it is
tempting to conclude that weak electroweak symmetry breaking is the
preferred mechanism, leading to an expected mass of the lightest Higgs
boson below 200~GeV (less than 135~GeV in the simplest supersymmetric
models), and a spectrum of additional neutral and charged Higgs bosons
with masses up to of order 1~TeV.

\section{EWSB Physics at Present and Near-Future Hadron Colliders: \protect\\
Tevatron and LHC}

\subsection{Standard Model Higgs Boson}

\begin{figure}[t!]
\begin{center}
\resizebox{\textwidth}{3.5in}{
\scalebox{1.25}[1.99]{
\includegraphics*[20,245][530,580]{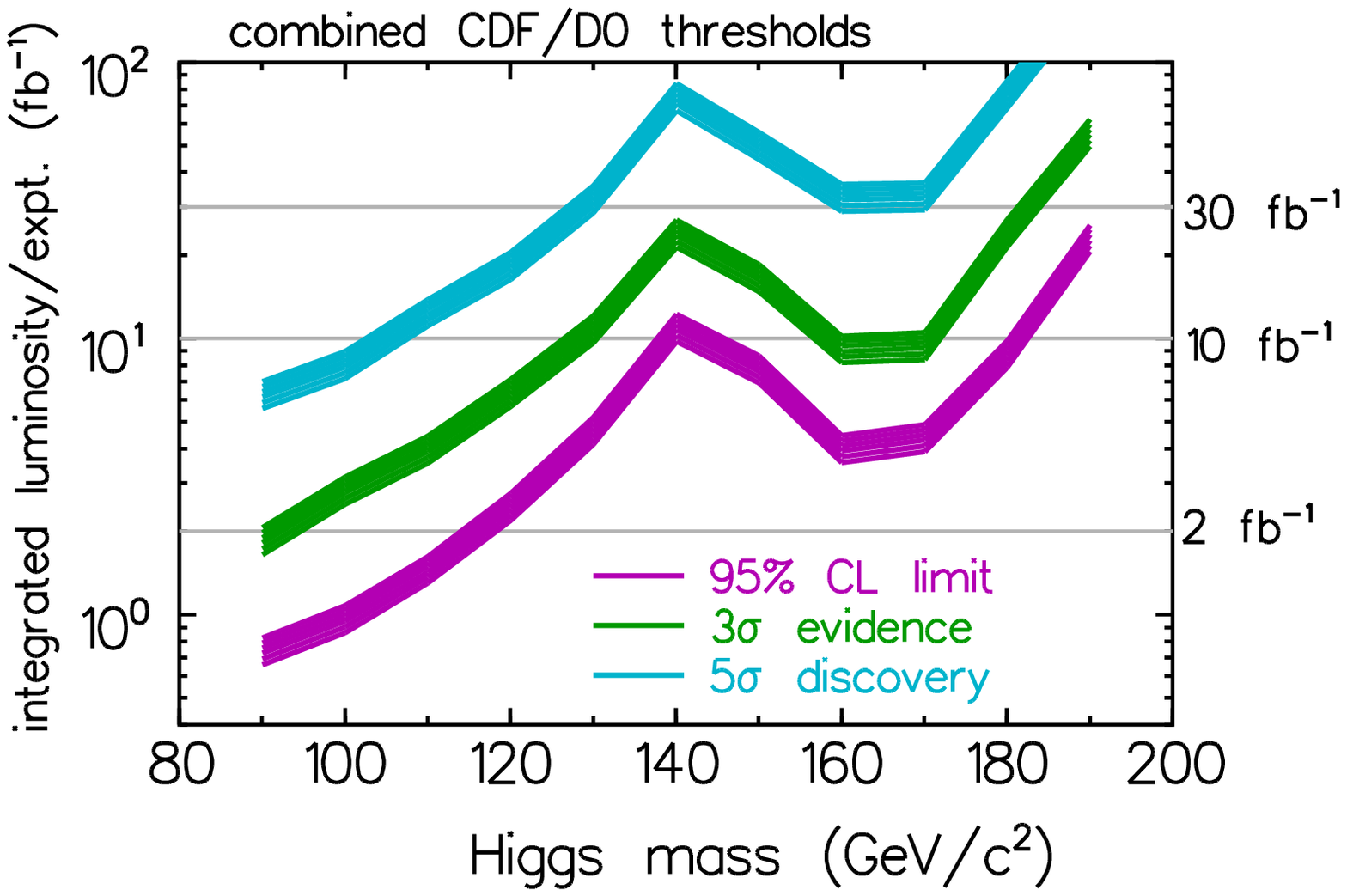}}
\includegraphics*[20,-17][567,567]{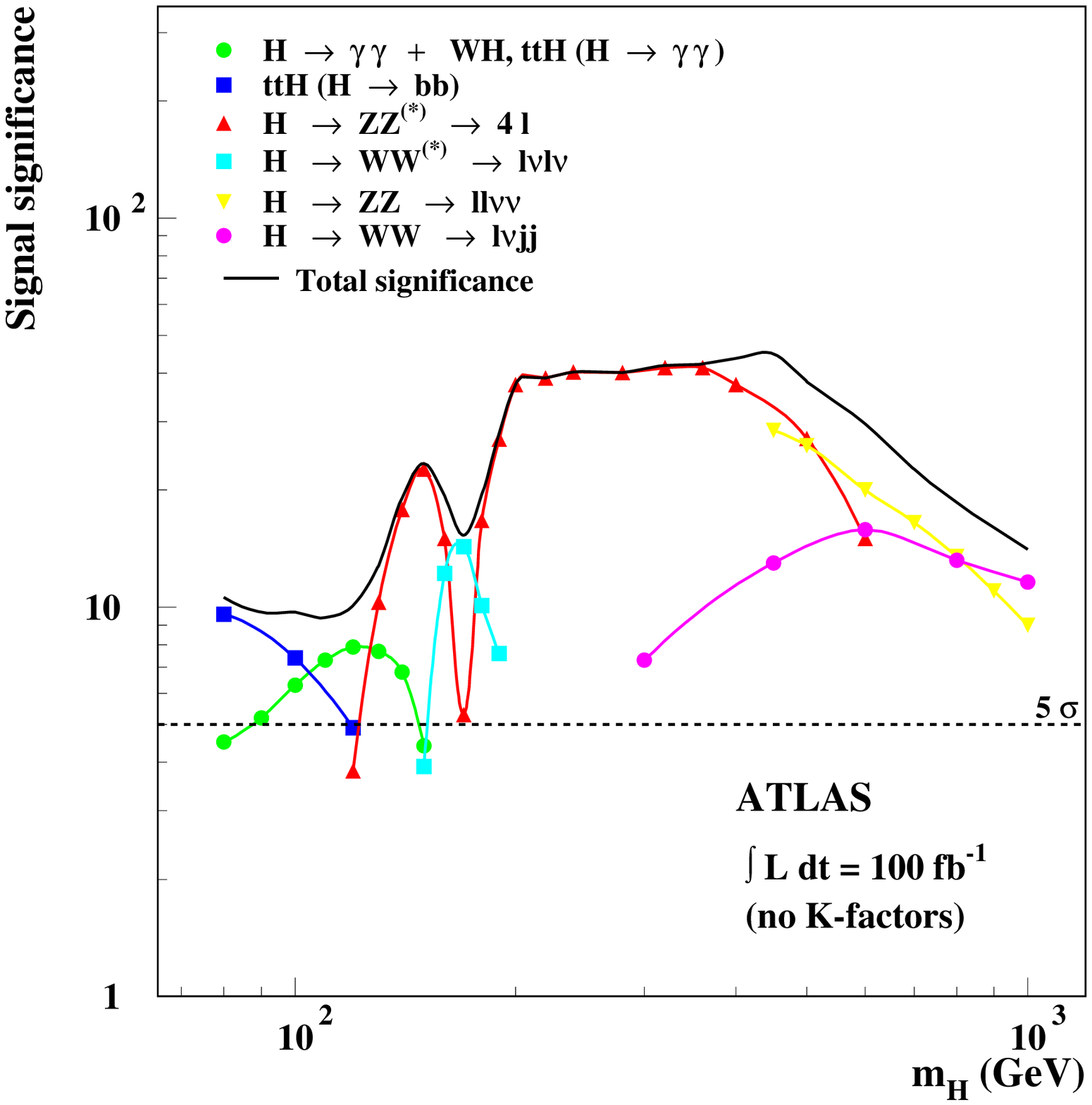}
}
\end{center}
\caption{(a)~The integrated luminosity required per Tevatron experiment, to
either exclude a Standard Model Higgs boson at 95\% CL or observe it at the
$3\sigma$ or $5\sigma$ level, as a function of the Higgs mass~\cite{Higgsrep}.
(b)~Higgs significance levels as a function of the 
Higgs mass for the ATLAS experiment at the LHC,
assuming an integrated luminosity of 
100~fb$^{-1}$~\cite{LHCreps}.}  
\label{fig:smhiggsathadron}
\end{figure}

After a decade long search for the Standard Model
(SM) Higgs boson ($\hsm$) at LEP,
Higgs masses up to 114 GeV have been excluded~\cite{lephiggs}.
The next step in the search for Higgs bosons will take place at
the Tevatron~\cite{Higgsrep}.  In the Higgs mass range below 135 GeV, 
the most promising signals can be extracted from $W\hsm$ and $Z\hsm$
Higgs-strahlung, in which the gauge bosons decay leptonically and the Higgs
boson decays into the $b\bar b$ final state.  For Higgs masses above 135 GeV, 
$\hsm\to WW^{(*)}$ becomes the dominant decay mode (the asterisk indicates
a virtual $W$).  The anticipated Tevatron Higgs discovery reach is
illustrated in Figure~\ref{fig:smhiggsathadron}a, and is
based on the combined statistical power of the CDF and D\O\
experiments~\cite{Higgsrep}.  
The curves shown are obtained by combining the $\ell\nu
b\bar b$, $\nu\bar\nu b\bar b$ and $\ell^+\ell^-b\bar b$ channels using
a neural network selection~\cite{neutralnet} in the low-mass Higgs region
($90~{\rm GeV}\lsim \mhsm\lsim 130$~GeV), and the
$\ell^\pm\ell^\pm jjX$ and $\ell^+\ell^-\nu\bar\nu$
channels~\cite{turcot} in the
high-mass Higgs region ($130~{\rm GeV}\lsim \mhsm\lsim 190$~GeV).  
The lower edge of the bands is the calculated threshold;
the bands extend upward from these nominal thresholds by 30\% as an
indication of the uncertainties in $b$-tagging efficiency, background
rate, mass resolution, and other effects.
Combining all the indicated channels,
the integrated luminosities necessary to rule out the Higgs boson of
the Standard Model for a mass below 200~GeV at the 95\% CL limit, or
to establish the observation of the Higgs boson at the 3$\sigma$ or $5\sigma$
level are displayed in Figure~\ref{fig:smhiggsathadron}a.  
Evidently, large integrated luminosities (10 to 30 fb$^{-1}$) are
needed to reach a definite conclusion on the observation
of the Higgs boson at the Tevatron.

Production rates for the Higgs boson in the Standard Model are
significantly larger at the LHC. The dominant Higgs
production process, gluon fusion, can be exploited in conjunction with
a variety of other channels, {\it e.g.}, $WW/ZZ$ fusion of the Higgs
boson and Higgs radiation off top quarks~\cite{LHCreps,branson,leshouches}. 
Integrated luminosities between 30 and 100
fb$^{-1}$, achievable within the first few years of LHC
operation, will be
sufficient to cover the entire canonical Higgs mass range of the
Standard Model up to values close to 1 TeV with a significance greater
than $5\sigma$
as shown in Figure~\ref{fig:smhiggsathadron}b.  Thus, there is no
escape route for the SM Higgs boson at the LHC.

If a SM Higgs boson is discovered at the Tevatron, the
Higgs mass can be measured with an accuracy of order 
2~GeV~\cite{precisiongroup}, whereas
the determination of Higgs couplings to $W$ and $Z$ bosons and to
bottom quarks will be model-dependent and fairly crude.  More precise
measurements of the properties of the Higgs boson mass in the Standard
Model can be performed at the LHC.  The $\hsm\to ZZ^{(*)}\to
\ell^+\ell^-\ell^+\ell^-$ channel allows for an accurate Higgs mass
determination of about $0.1\%$ for $120~{\rm GeV} \lsim\mhsm\lsim
400$~GeV, assuming an integrated luminosity of 
300~fb$^{-1}$~\cite{trefzger}.  
For larger Higgs masses, the precision in the Higgs
mass measurement deteriorates due to the effect of the increasing
Higgs width; nevertheless a $1\%$ Higgs mass measurement is possible
for $\mhsm\simeq 700$~GeV.  The Higgs width can be extracted with a
precision of 5 to $6\%$ over the mass range $300$---700~GeV from the
Breit-Wigner shape of the Higgs resonance~\cite{trefzger}.  Below
300~GeV, the instrumental resolution becomes larger than the Higgs width,
and the accuracy of the Higgs width measurement degrades.  For
example, the four-lepton invariant mass spectrum from $\hsm\to ZZ$ yields a
precision of about $25\%$ at $\mhsm=240$~GeV~\cite{precisiongroup}.
For lower Higgs masses, indirect methods 
must be employed to
measure the Higgs width.

For Higgs masses below 200 GeV, a number of different Higgs decay
channels can be studied at the LHC.  
The relevant processes are~\cite{branson,xseca}:
\beqa
&& gg\to\hsm\to\gamma\gamma\,, \nonumber \\
&& gg\to\hsm\to VV^{(*)}\,, \nonumber \\
&& qq\to qqV^{(*)} V^{(*)}\to qq\hsm,\quad \hsm\to\gamma\gamma,\,
\tau^+\tau^-, \,VV^{(*)}\,, \nonumber \\
&& gg, q\bar q\to t\bar t\hsm, \quad \hsm\to b\bar b,
\,\gamma\gamma, \,WW^{(*)}\,,\nonumber
\eeqa
where $V=W$ or $Z$.
The gluon-gluon fusion mechanism is the dominant Higgs production
mechanism at the LHC, yielding a total cross section of about 30~pb
[15~pb] for $\mhsm=120$~GeV [$\mhsm=200$~GeV].
One also has
appreciable Higgs production via $VV$ electroweak gauge boson fusion,
with a total cross section of about 6~pb
[3~pb] for the Higgs masses quoted above.  
The electroweak gauge boson fusion mechanism can be separated from the
gluon fusion process by employing a forward jet tag
and central jet vetoing techniques.
The cross section for $t\bar t\hsm$ production can be significant for
Higgs masses in the intermediate mass range~\cite{xsecb},
0.8~pb [0.2~pb] at $\mhsm=120$~GeV
[$\mhsm=200$~GeV], although this cross section falls faster with Higgs
mass as compared to the gluon and gauge boson fusion mechanisms.

\begin{figure}[t!]
\begin{center}
\resizebox{.85\textwidth}{!}{
\includegraphics*[angle=90]{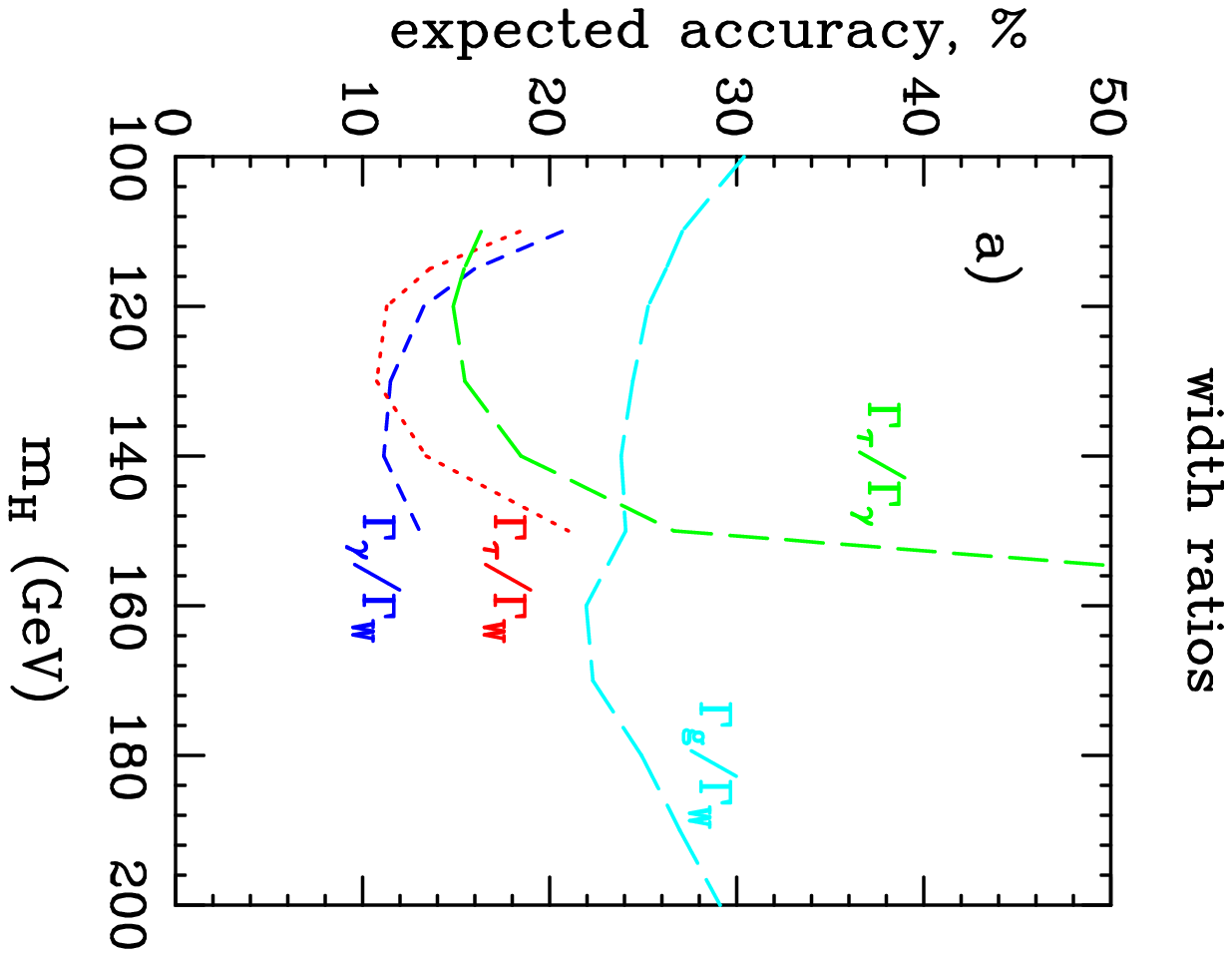}
\hspace*{3mm}
\includegraphics*[angle=90]{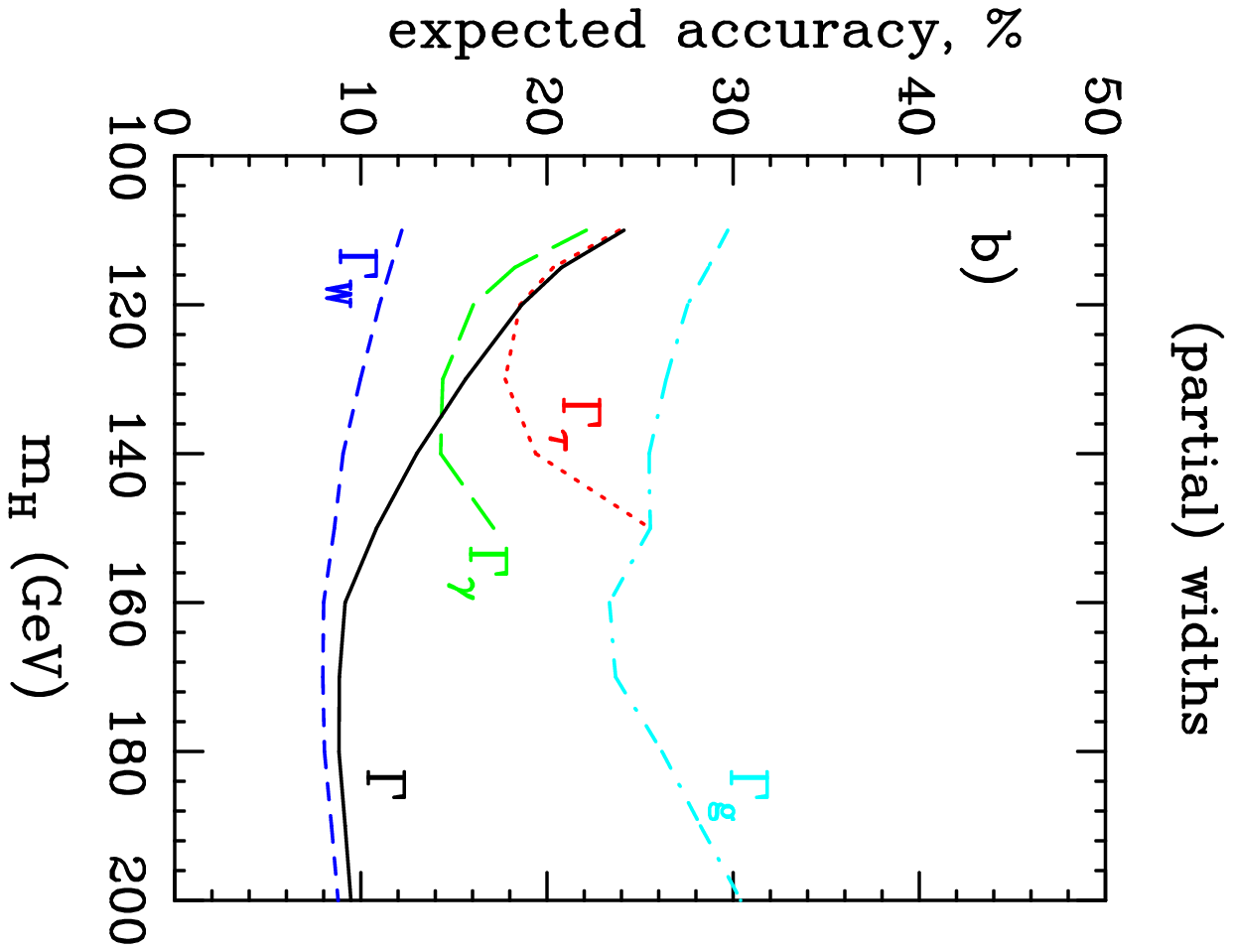}
}
\end{center}
\caption[0]{Relative accuracy expected at the LHC with 
200~fb${^{-1}}$ of data for (a)~various ratios of Higgs boson partial
widths and (b)~the indirect determination of partial and total
widths.  Expectations for width
ratios assume $W$, $Z$ universality;
indirect width measurements also assume $b$, $\tau$ universality and
a small branching ratio for unobserved modes.  Taken from the parton-level
analysis of Ref.~\protect\cite{zeppenfeld}.}
\label{fig:LHCwidths}
\end{figure}

The measurements of
various relations between Higgs decay branching ratios can 
be used to infer the
ratios of various Higgs couplings, and provide 
an important first step in clarifying the
nature of the Higgs boson.  These can be extracted 
from a variety of Higgs signals which are observable over a limited
range of Higgs masses.
In the mass range $110~{\rm GeV}\lsim\mhsm 
\lsim 150$~GeV, the Higgs boson can be
detected [with 100 fb$^{-1}$ of data]
in the $\gamma\gamma$ and the $\tau^+\tau^-$ channels
indicated above.
For $\mhsm\gsim 130$~GeV, the Higgs boson can also be detected
in gluon-gluon fusion through its decay to $WW^{(*)}$, 
with both final gauge bosons decaying leptonically~\cite{dreiner}, 
and to $ZZ^{(*)}$
in the four-lepton decay mode~\cite{LHCreps,branson}.
There is additional
sensitivity to Higgs production via $VV$ fusion followed by its decay
to $WW^{(*)}$ for $\mhsm\gsim 120$~GeV.  
These data can be used to extract
the ratios of the Higgs partial widths to gluon pairs,
photon pairs, $\tau^+\tau^-$,
and $W^+W^-$~\cite{zeppenfeld,PlehnRainwaterZeppenfeld}.
The expected accuracies in Higgs width
ratios, partial widths, and the total Higgs width 
are exhibited in Figure~\ref{fig:LHCwidths}. 
These results are obtained under the assumption that the
partial Higgs widths to $W^+W^-$ and $ZZ$ are fixed by electroweak
gauge invariance, and the ratio of the
partial Higgs widths to $b\bar b$ and $\tau^+\tau^-$ are fixed by 
the universality of Higgs couplings to down-type fermions.
One can then extract the total Higgs width under the assumption
that all other unobserved modes, in the Standard Model and beyond,
possess small branching ratios of order 1\%.
Finally, we note that 
the specific Lorentz structure predicted for the $\hsm W^+W^-$
coupling by the Higgs
mechanism can be tested in angular correlations between the spectator
jets in $WW$ fusion of the Higgs boson at the 
LHC~\cite{PlehnRainwaterZeppenfeld}.

With an integrated luminosity of 100 fb$^{-1}$ per
experiment, the relative accuracy expected at the LHC for
various ratios of Higgs partial widths $\Gamma_i$
range from 10\% to 30\%, as shown in
Figure~\ref{fig:LHCwidths}. These
correspond to 5\% to 15\% measurements of various ratios of Higgs
couplings.  The ratio $\Gamma_\tau/\Gamma_W$ measures the coupling of
down-type fermions relative to the Higgs couplings to gauge bosons. 
To the extent that the one-loop $\hsm\gamma\gamma$ amplitude is
dominated by the $W$-loop, the partial width ratio
$\Gamma_\tau/\Gamma_\gamma$ probes the same relationship.  In
contrast, under the usual assumption that the one-loop $\hsm gg$ amplitude 
is dominated by the top-quark loop, the ratio $\Gamma_g/\Gamma_W$
probes the coupling of up-type fermions relative to the $\hsm WW$
coupling.  Additional information about Higgs couplings can be
ascertained by making use of the $t\bar t\hsm$ production mode at the LHC,
followed by $\hsm\to b\bar b$.  Recent 
studies~\cite{ATLASstudy,CMSstudy} by the ATLAS
and CMS collaborations suggest that for an integrated luminosity of 
100~fb$^{-1}$, this signal is viable 
if $\mhsm\lsim 130$~GeV.
Including the $t\bar t\hsm$ mode allows for an independent
check of the Higgs-top quark Yukawa coupling.
Moreover, if
combined with information obtained from $\Gamma_g$, one can
test, through the decay $\hsm\to b\bar b$,
the assumption of universality of Higgs
couplings to down-type fermions.  

Finally, one can check the consistency of the Standard Model by
comparing the observed Higgs mass to the value deduced from precision
electroweak fits.  With improvements expected both for the precision
in the measured values of $m_W$, $m_t$ and the electroweak mixing angle, one
can anticipate an improvement in the fractional $1\sigma$ uncertainty
in the Higgs mass at future colliders~\cite{futureprecision}.  
After 2~fb$^{-1}$ [15~fb$^{-1}$] of integrated
luminosity at the Tevatron, the anticipated fractional Higgs mass
uncertainty will decrease to about $35\%$ [$25\%$].  Further
improvements at the LHC with 100~fb$^{-1}$ of data can reduce this
uncertainty to about $18\%$.  This will yield strong constraints on
the Standard Model and could provide evidence for new physics if a
disagreement is found between the inferred Higgs mass from precision
measurements and the actual Higgs mass, or if no Higgs boson 
is discovered.

\subsection{Higgs Bosons in Supersymmetric Extensions of the Standard Model}

In supersymmetric extensions of the Standard Model, there is one
neutral Higgs state which often exhibits properties similar to
those of the SM Higgs boson.  In addition, new neutral and
charged scalar states arise whose properties encode the physics of the
electroweak symmetry breaking dynamics.  In the absence of CP
violation, the neutral Higgs bosons carry definite CP quantum numbers.

\begin{figure}[b!]
\begin{center}
\resizebox{\textwidth}{!}{
\includegraphics*[0,0][567,567]{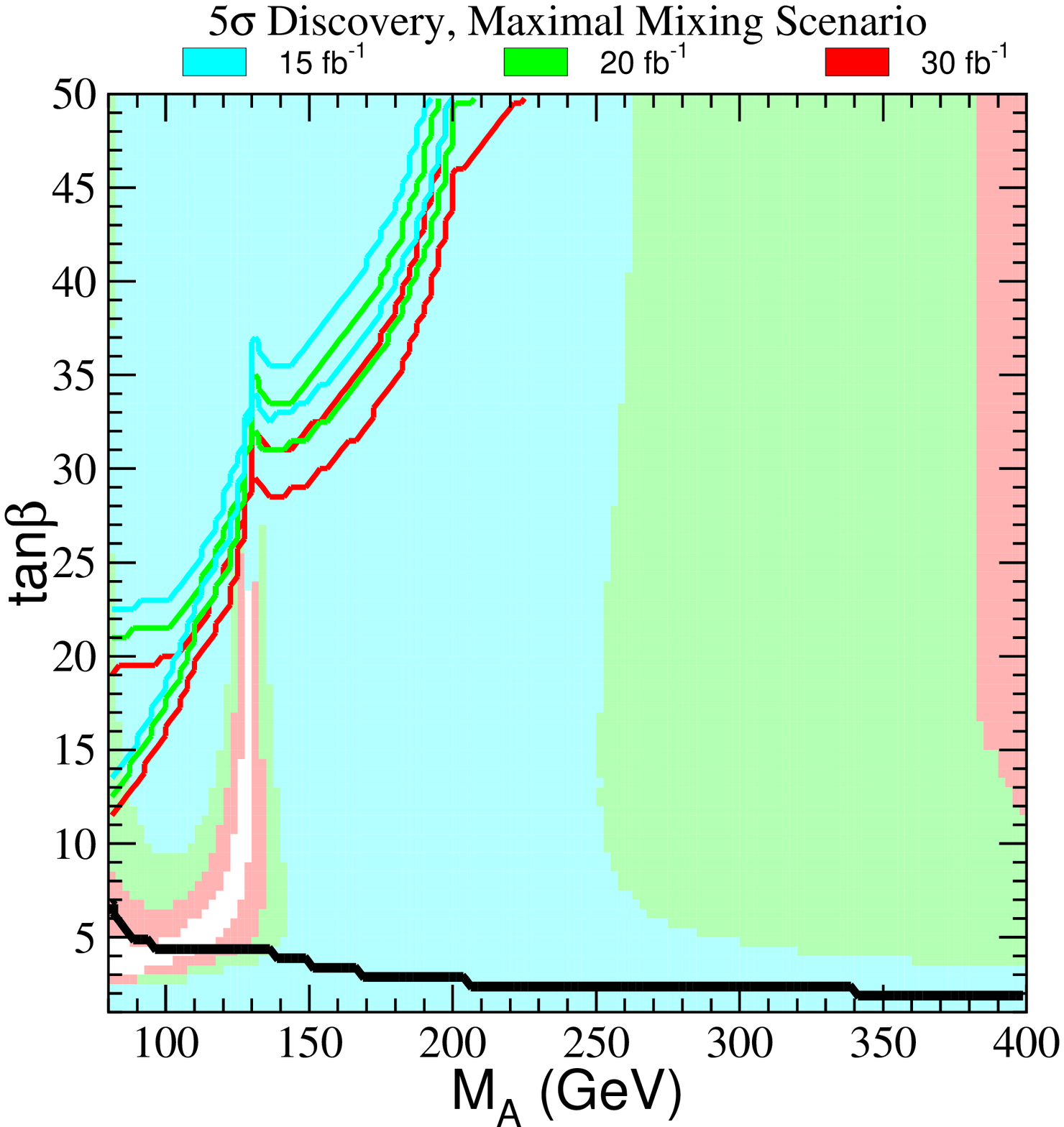}
\includegraphics*[0,0][530,520]{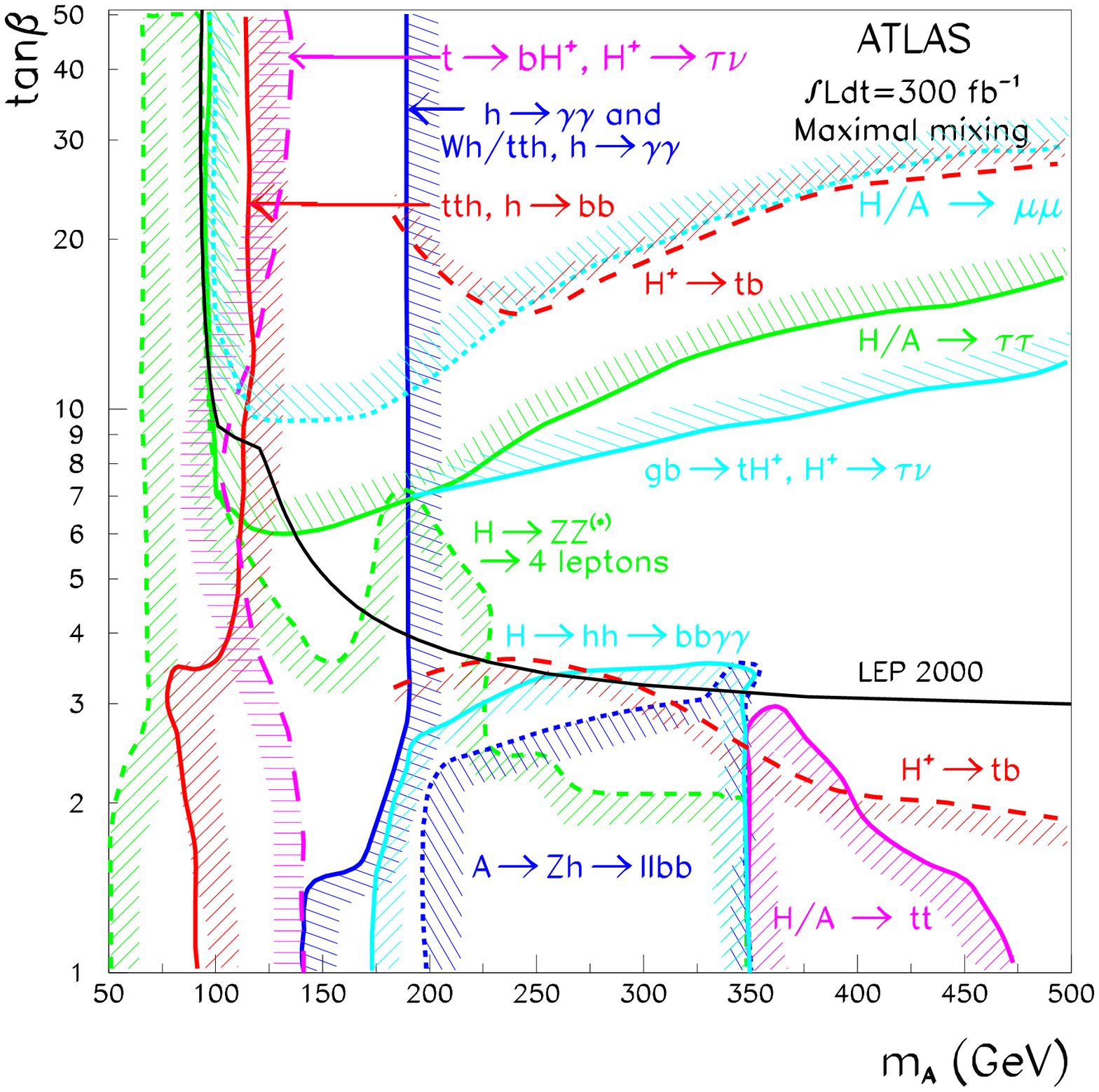}
}
\end{center}
\caption{(a)~$5\sigma$ discovery region on the $m_A$--$\tan \beta$
plane~\cite{Higgsrep}, 
for the maximal mixing scenario and two different search channels:
$q\bar q\to V\phi$ ($\phi=h$, $H$), $\phi\to b\bar b$
(shaded regions) and $gg$, $q\bar q\to b\bar b\phi$ ($\phi=h$,
$H$, $A$), $\phi\to b\bar b$ (region in the upper left-hand corner
bounded by the solid lines).  Different integrated luminosities are
explicitly shown by the color coding.  The two sets of lines (for a
given color) correspond to the CDF and D\O\ simulations, respectively.
The region below the solid black line near the bottom of the plot is
excluded by the absence of observed $e^+e^-\to Z\phi$ events at LEP2.
(b)~$5\sigma$ discovery contours for MSSM Higgs boson detection in
various channels are shown in the $m_A$--$\tan\beta$ parameter space,
in the maximal mixing scenario, assuming 
an integrated luminosity of $L=300~{\rm fb}^{-1}$
for the ATLAS detector~\cite{LHCreps,branson}. } 
\label{fig:susyhiggsathadron}
\end{figure}

In the minimal supersymmetric extension of the Standard
Model (MSSM), the tree-level Higgs sector is automatically CP conserving.  
CP-violating effects can enter via loop corrections, and can be
significant in certain regions of MSSM parameter space.  However,
unless otherwise noted, we will neglect CP-violating Higgs sector
effects in what follows. 
The mass of the lightest CP-even neutral Higgs boson ($h$) of the MSSM
is less than about 135 GeV~\cite{higgslim}.  
This prediction incorporates significant
radiative corrections, which shift the maximal Higgs mass from
its tree-level value of $m_Z$~\cite{hhprl}.  
This maximal mass is achieved when
the top-squark mixing parameters are such that
the contribution from
the radiative corrections associated with loops of top-squarks is
maximal (this is the {\it maximal mixing} scenario).
In addition, the Higgs spectrum contains a
heavier CP-even neutral Higgs boson ($H$), a CP-odd neutral Higgs
boson ($A$) and a charged Higgs pair ($H^\pm$).  In contrast
to the $h$ mass, the masses of the $H$, $A$ and $H^\pm$ Higgs bosons
are not similarly constrained and
can be significantly larger than the $Z$ mass.
In the MSSM, the tree-level Higgs sector is fixed by the values of
$m_A$ and the ratio of Higgs vacuum expectation values, $\tan\beta$.  
When radiative corrections are included, additional MSSM parameters
enter and determine the size of the loop corrections.  
For example, in the maximal mixing scenario
(with other MSSM parameters specified according to Table~53 of
Ref.~\cite{Higgsrep}), most
of the $m_A$--$\tan\beta$ parameter space can be covered at the
Tevatron given sufficient luminosity
[shaded areas in Figure~\ref{fig:susyhiggsathadron}a]
by the search for 
CP-even Higgs bosons with significant couplings to the $W$ and $Z$.
The remaining unexplored regions will be 
covered at the LHC [Figure~\ref{fig:susyhiggsathadron}b].  That is,
at least one of the Higgs bosons of the MSSM
is guaranteed to be discovered at either the Tevatron and/or the LHC.
The coverage in the $m_A$--$\tan\beta$ plane
by different Higgs production and decay channels can
vary significantly, depending on the choice of MSSM parameters.
For example, if the CP-even Higgs boson with the larger
coupling to the $W$ and $Z$  has a strongly suppressed coupling to 
bottom quarks, the Higgs searches at the Tevatron will become 
more problematical, while the LHC search for Higgs production followed
by its decay into photons becomes more favorable~\cite{carenaetal}.

In some regions of MSSM parameter space, more than one Higgs boson
can be discovered at the LHC.  However, there is a sizable wedge-shaped 
region of the parameter space at moderate values of $\tan\beta$
opening up from about $m_A=200$~GeV to higher values in which
the heavier Higgs bosons cannot
be discovered at the LHC 
[see Figure~\ref{fig:susyhiggsathadron}b].  
In this region of the MSSM parameter space,
only the lightest CP-even Higgs boson can be discovered, and its properties
are nearly indistinguishable from those of the SM Higgs
boson.
Deviations from SM properties can also occur if the
Higgs decay into supersymmetric particles is kinematically allowed,
or if light supersymmetric particles contribute significantly to
Higgs loop amplitudes.
High precision measurements of Higgs branching ratios and other properties
will be required in order to detect deviations from 
SM Higgs predictions and
demonstrate the existence of a non-minimal Higgs sector.

The phenomenology of the MSSM Higgs sector is closely tied to various
MSSM parameters that arise as a consequence of low-energy
supersymmetry breaking.  A priori, one can parameterize this breaking
in terms of the most general set of soft supersymmetry-breaking
terms~\cite{susybreaking}.  Alternatively, one can propose models of
fundamental supersymmetry breaking, which constrain many of these
terms.  The {\it Snowmass points and slopes (SPS)}, developed in
Ref.~\cite{SPS}, are a consensus of benchmark points and model lines
(``slopes'') within various theoretical approaches to supersymmetry
breaking, which correspond to different scenarios in the search for
supersymmetry at future colliders.  One expects a significant
interplay between the study of supersymmetric phenomena and the
observation of Higgs bosons and their properties.  Ultimately, one
hopes to learn if (and how) the origin of electroweak breaking depends
fundamentally on the physics of supersymmetry breaking.

\subsection{Strong Electroweak Symmetry Breaking Dynamics}

If strong electroweak symmetry breaking with no Higgs boson in the
mass range below 1 TeV is realized in nature, the Tevatron may provide
the first hints of new physics, while LHC can provide some
insight into the domain of the new strong interactions~\cite{han}.  
The top quark
may play a critical role in this enterprise, due to the fact that its
large mass implies the strongest coupling to the
electroweak symmetry breaking sector, compared to the other known
particles of the Standard Model.  At the Tevatron, hints of new
physics associated with the top quark can emerge in a number of ways.  
Anomalous top quark production and/or new particles that decay into
$t\bar t$ pairs would be a possible signal of strong electroweak
symmetry-breaking dynamics.

At the LHC,
deviations from the perturbative predictions for $W^+W^-$
production in quark-antiquark collisions shed light on the onset of
the new interactions between the $W$ bosons below 3 TeV. This
range is also expected to be covered in strong $WW$ quasi-elastic
scattering. Access to this new domain can also be provided by
observing pseudo-Goldstone bosons associated with the spontaneous
breaking of global symmetries of the new strong 
interactions~\cite{casalbuoni}.  
In addition, the
observation of genuine new resonances (made up of techniquarks or other
new fundamental strongly-interacting particles)
is possible for masses below 2 to 3 TeV~\cite{branson}.  
Evidence for new substructure can also be detected indirectly
via deviations in jet-jet and Drell-Yan cross sections.  For example,
with 300 fb$^{-1}$ of data, the measurement of the dijet cross section
is sensitive to a compositeness scale of about 40~TeV.
However, the energy of the LHC and the resolution of the
experiments fall short of a detailed analysis of the new strong
interactions.

\section{EWSB Physics at Future $e^+ e^-$ Linear Colliders} 

\subsection{Standard Model Higgs Boson}

The next generation of high energy $e^+e^-$ linear colliders
is expected to operate 
at energies from 300 GeV up to about 1 TeV (JLC, NLC, TESLA),
henceforth referred to as the LC~\cite{jlc,nlc,tesla}.
The possibility of a multi-TeV linear collider 
operating in an energy range of 3--5 TeV (CLIC) is also under 
study~\cite{clic}.
With the expected high luminosities, up to 1~ab$^{-1}$,
accumulated within a few years in a clean experimental environment,
these colliders are ideal instruments for reconstructing the
mechanism of electroweak symmetry breaking in a comprehensive and
conclusive form.

If weakly-coupled electroweak symmetry breaking dynamics
(involving an elementary scalar Higgs field) is
realized in nature, then it can be established experimentally
in three steps:

\begin{enumerate}
\item
The Higgs boson must be observed clearly and unambiguously, and its
basic properties---mass, width, spin and C and P quantum numbers---must be
determined.

\item
The couplings of the Higgs boson to the $W^\pm$ and $Z$ 
bosons and to leptons and quarks must be measured.  Demonstrating that
these couplings scale with the mass of the corresponding
particle would provide a critical verification
of the Higgs mechanism as the responsible agent
for generating the masses of the fundamental particles.

\item
The Higgs potential must be reconstructed by measuring the self-coupling
of the Higgs field. The specific form of the potential shifts the ground
state to a non-zero value, thereby providing the mechanism for
electroweak symmetry breaking based on the self-interactions of scalar
fields. 
\end{enumerate}

\begin{figure}[t!]
\begin{center}
\resizebox{\textwidth}{!}{
\includegraphics{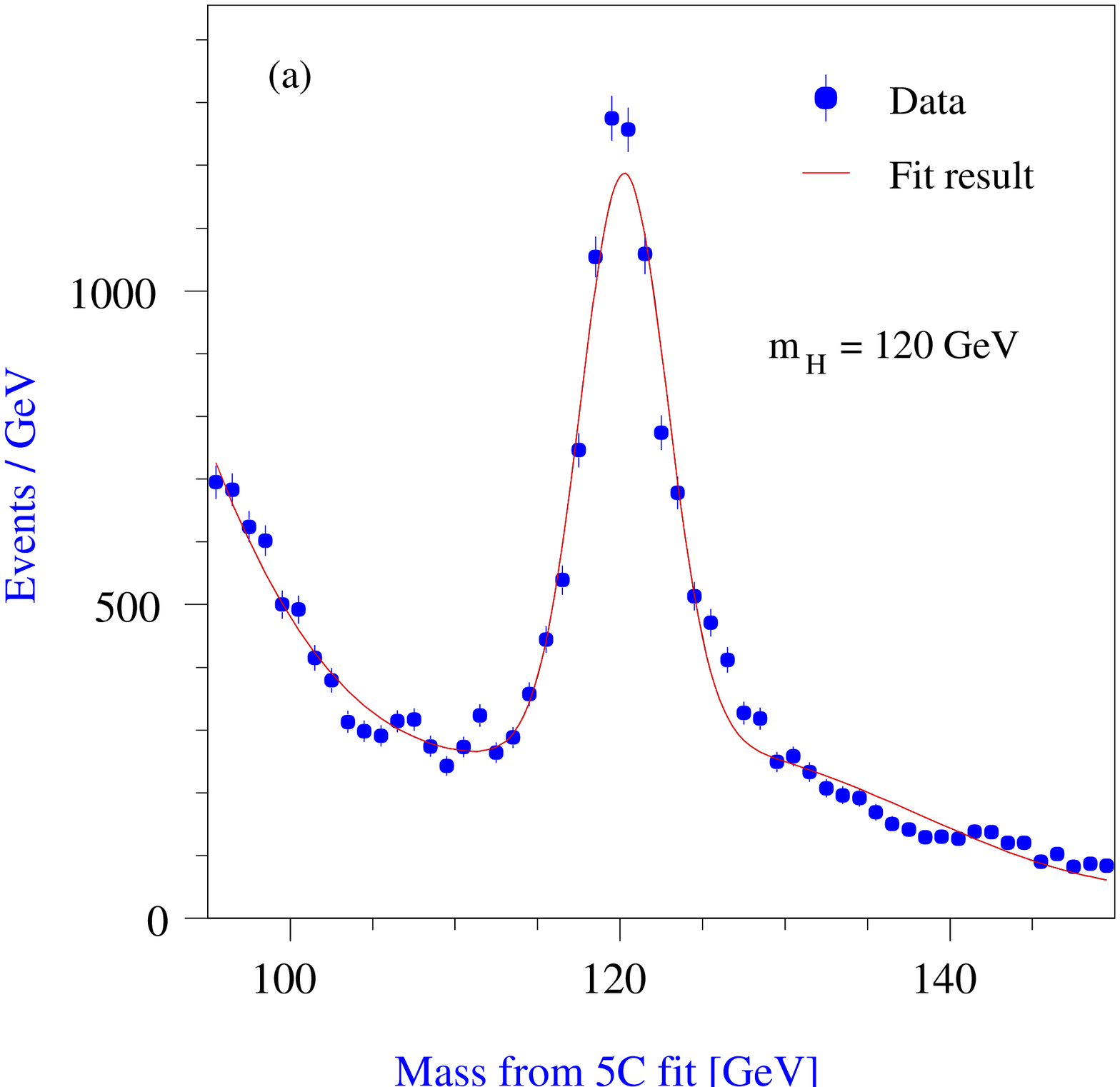}
\hspace*{3mm}
\includegraphics{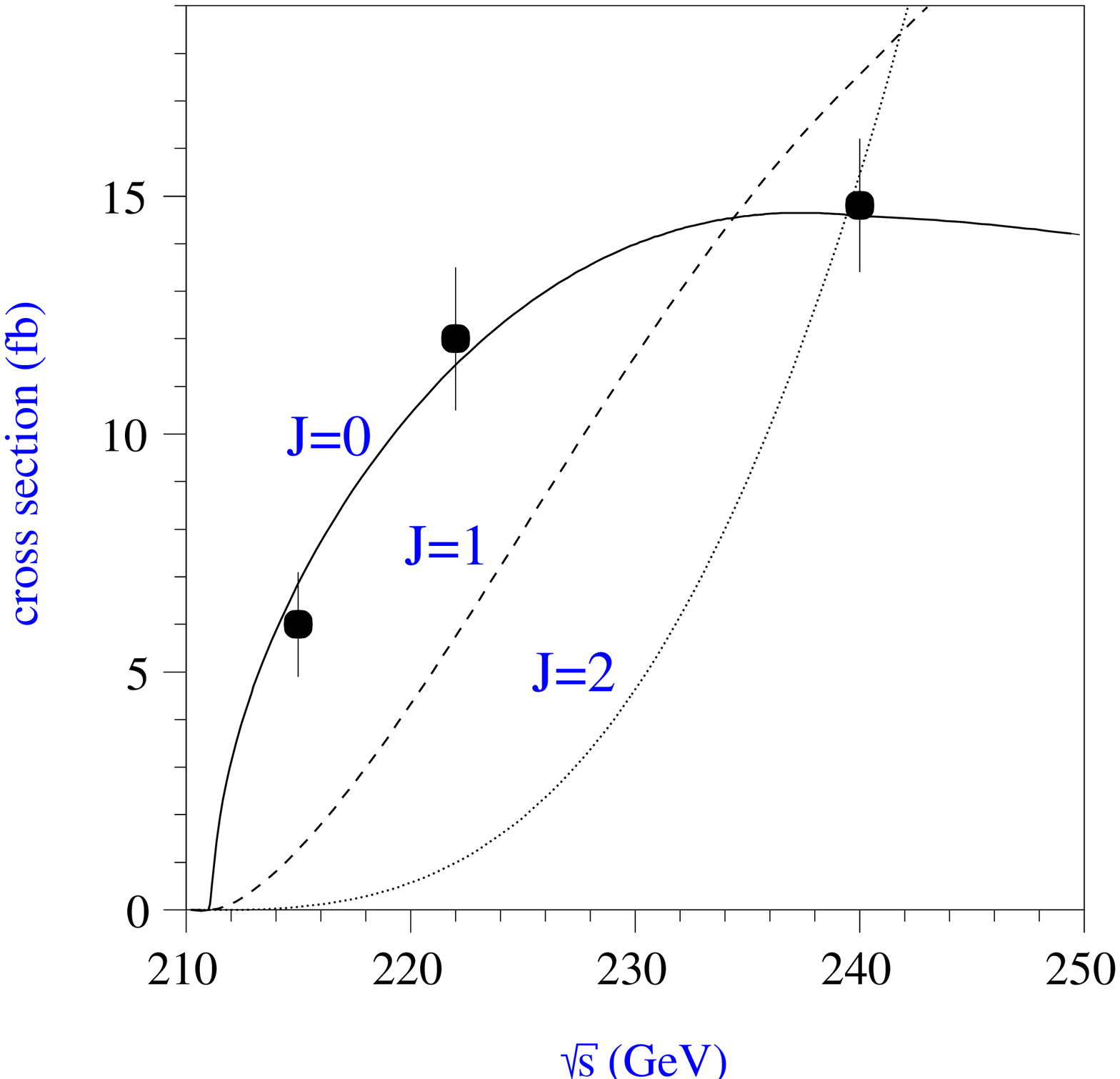}
}
\hspace*{7.3mm}
\includegraphics[width=2.923in,height=3.2in]{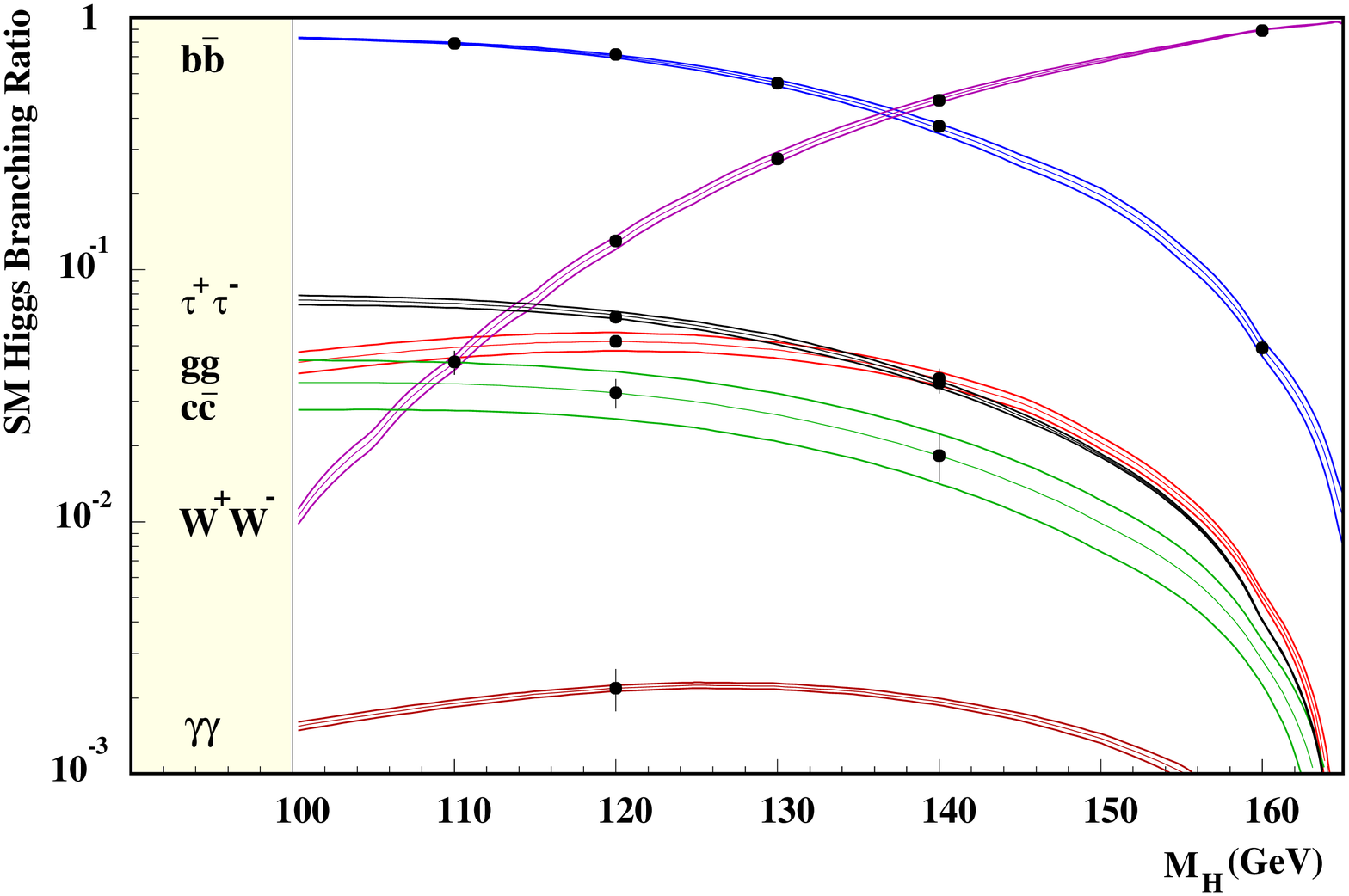}
\hspace*{10.7mm}\vspace*{-0.8mm}
\includegraphics[width=2.864in,height=3.17in]{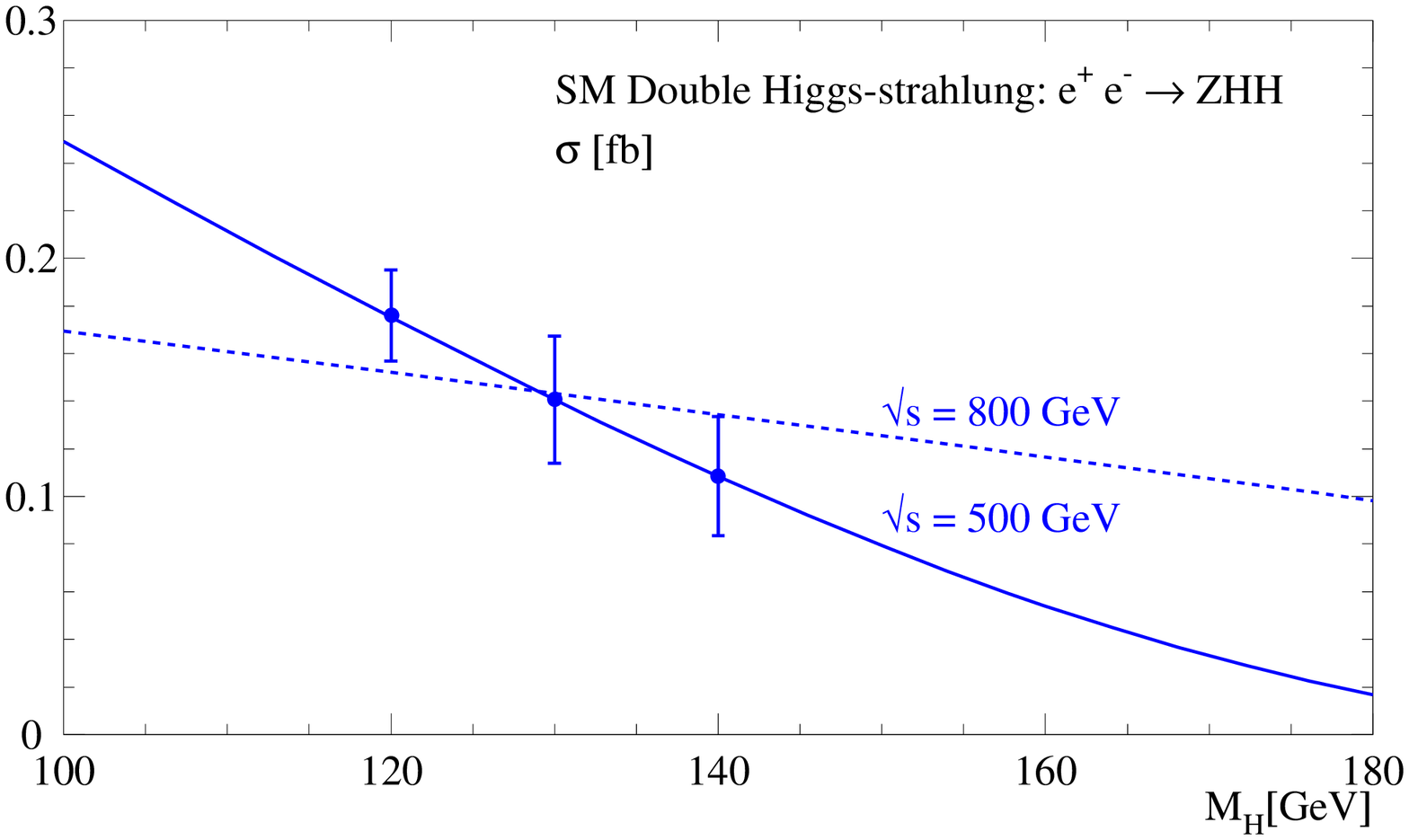}
\end{center}
        \caption{(a)~The Higgs boson mass peak reconstructed in
 the channel $e^+e^-\to Z\hsm \rightarrow b\bar{b} q \bar{q}$ for   
$\mhsm = 120$~GeV~\cite{Brau}; 
(b)~Simulated measurement of the 
 e$^+$ e$^- \rightarrow Z\hsm$ cross section  for $\mhsm = 120$~GeV  with
 20~fb$^{-1}$/point at three center of mass energies compared to the 
predictions for spin-0 (full line) and typical examples of spin-1 
particles (dashed line) and spin-2 particles (dotted line)~\cite{hspin}; 
(c)~The predicted SM Higgs boson branching ratios.
 Points with error bars show the expected experimental accuracy, while the
 lines indicate the theoretical uncertainties on SM 
predictions~\cite{BattagliaDesch}.
(d)~Cross section for the double Higgs-strahlung process $e^+e^-\to 
Z\hsm\hsm$ at
$\sqrt{s}=500$~GeV (solid line) and 800~GeV (dashed line)~\cite{n36a}. 
 The data points show the accuracy for 1~ab$^{-1}$.}
        \label{fig:LChiggsproperties}
\end{figure}

Essential elements of this program can be realized at a high-luminosity
$e^+e^-$ linear collider~\cite{Tesla-TDR,LC-Orange,JLC-TDR}.
With an accumulated luminosity of 500 fb$^{-1}$, about $10^5$ Higgs
bosons can be produced by Higgs-strahlung 
$e^+e^- \rightarrow Z\hsm$
in the theoretically
preferred intermediate mass range below 200 GeV.  Given the low background,
as illustrated in 
Figure~\ref{fig:LChiggsproperties}a~\cite{Brau}, high-precision
analyses of the Higgs boson are possible in these machines.
The Higgs mass will be measured to an accuracy of order 100 MeV
(with an achievable fractional precision of
$5\times 10^{-4}$ for $\mhsm=120$~GeV). The Higgs width 
can be inferred, in a model-independent way, to an accuracy up to 5~\%,
by combining
the partial width to $W^+W^-$, accessible in the vector boson fusion process,
with the $W^+W^-$ decay branching ratio. Spin and parity can be determined
unambiguously from the steep onset of the excitation curve in
Higgs-strahlung near the threshold 
(see Figure~\ref{fig:LChiggsproperties}b~\cite{hspin}) 
and the angular correlations in this process.

Higgs decay branching ratios can be measured very precisely in the
intermediate mass range 
(as shown in Figure~\ref{fig:LChiggsproperties}c~\cite{BattagliaDesch}).
When these measurements are combined with measurements of Higgs
production cross sections, the absolute values of the Higgs couplings
to the $W^\pm$ and $Z$ gauge bosons and the Yukawa couplings to
leptons and quarks can be determined to a few percent in a
model-independent way.  In addition, the Higgs-top Yukawa coupling can 
be inferred from the cross section for Higgs emission off 
top-antitop quark pairs~\cite{tth1,tth2}.  The exected accuracies for the
measurements of Higgs couplings is given in Table~\ref{tab:BRmeas}. These
observations are essential for establishing weakly-coupled scalar
dynamics and the associated Yukawa interactions as the
mechanism generating the masses of the fundamental particles in the
Standard Model.

\begin{table}[t!]
        \begin{center} \begin{tabular}{|c||c|c|}
\hline
           Coupling & $\mhsm = 120$~GeV & $\mhsm = 140$~GeV \\ \hline\hline
           $\hsm WW$    &      1.3\%     &  2.0\% \\
           $\hsm ZZ$    &      1.2\%     &  1.3\% \\ \hline
        $\hsm t\bar{t}$ &      3.0\%     &  6.1\% \\
        $\hsm b\bar{b}$ &      2.2\%     &  2.2\% \\
        $\hsm c\bar{c}$ &      3.7\%     & 10.2\% \\ \hline
        $\hsm \tau\tau$ &      3.3\%     &  4.8\% \\ \hline
        \end{tabular} \end{center}
\caption{Expected accuracies for measurements of Higgs
couplings at an $e^+e^-$ linear collider for Higgs masses of 120
and 140~GeV in the Standard Model, from
Ref.~\protect\cite{BattagliaDesch}. For the $WW$ and $ZZ$
couplings, 500~fb$^{-1}$ at $\sqrt{s}=500$~GeV are assumed. For
$b\bar b$, $c\bar c$, and $\tau\tau$, the study assumes
500~fb$^{-1}$ at $\sqrt{s}=350~{\rm GeV}$; for $t\bar t$,
1~ab$^{-1}$ at $\sqrt{s}=800$~GeV.}  \label{tab:BRmeas}
\end{table}

The measurement of the self-couplings of the Higgs field is a very
ambitious task that requires the highest luminosities possible at
$e^+e^-$ linear colliders, which possess unique capabilities for addressing
this question. The trilinear Higgs self-coupling can be measured in
double Higgs-strahlung, in which a virtual Higgs boson splits into two
real Higgs particles in the final state~\cite{n36a,trilinear}.
A simulation based on 1~ab$^{-1}$ of data is
exhibited in Figure~\ref{fig:LChiggsproperties}d~\cite{n36a}.
In this way, the cubic term of the scalar potential
can be established at a precision of about 20\%.
Such a measurement is a prerequisite for developing the form of the
Higgs potential specific for spontaneous electroweak symmetry breaking in the
scalar sector.

If the SM Higgs mass is above 200~GeV, then the precision
determination of Higgs couplings will have to be reconsidered.  The SM
Higgs discovery reach at the LC is maximized by 
considering both the Higgs-strahlung process, $e^+e^-\to Z\hsm$, and
the vector boson fusion process, $e^+e^-\to
\nu\bar\nu\hsm$.  For example, the analysis of Ref.~\cite{heavyHiggs}
suggests that for an integrated luminosity of 500~fb$^{-1}$, a Higgs
boson with mass up to about 650 GeV will be observable at the LC
with $\sqrt{s}=800$~GeV.  For Higgs masses above $t\bar t$ threshold,
one can measure the $t\bar t\hsm$ Yukawa coupling by observing Higgs
bosons produced by vector boson fusion which subsequently
decay to $t\bar t$.  The analysis of Ref.~\cite{topHiggs} finds that 
at the LC with $\sqrt{s}=800$~GeV and 1~ab$^{-1}$ of data,
the $t\bar t\hsm$ Yukawa coupling can be determined with an accuracy
of about 10\% for a Higgs mass in the range 350---500~GeV.

The $e^+e^-$ linear collider with center-of-mass energy $\sqrt{s}$
can also be designed to operate in a
$\gamma\gamma$ collision mode.  
This is achieved by using Compton 
backscattered photons in the scattering of intense laser 
photons on the initial polarized $e^\pm$ beams~\cite{n36b,boos}.
The resulting
$\gamma\gamma$ center of mass energy is peaked for proper
choices of machine parameters at about $0.8\sqrt{s}$.  The luminosity
achievable as a function of the photon beam energy depends strongly on
the machine parameters (in particular, the choice of laser
polarizations).  The photon collider provides additional opportunities 
for Higgs physics~\cite{hggpheno,higgsgamgam,boos,asner,velasco}.  
The Higgs boson can be produced as an
$s$-channel resonance in $\gamma\gamma$ collisions, and one can perform 
independent measurements of various Higgs couplings.  
For example, the product
$\Gamma(\hsm\to\gamma\gamma){\rm BR}(\hsm\to b\bar b)$ can be measured with a
statistical accuracy of about $2$---$10\%$ for 120~GeV$\lsim\mhsm\lsim 160$~GeV
with about 50~fb$^{-1}$ of data~\cite{higgsgamgam,asner,velasco}.  
Using values for BR$(\hsm\to b\bar b)$ and
BR$(\hsm\to\gamma\gamma)$ measured at the $e^+e^-$ linear collider,
one can obtain a value for the total Higgs width with
an error dominated by the expected error in BR$(\hsm\to\gamma\gamma)$.  
For heavier Higgs bosons, $\mhsm\gsim 200$~GeV, the total Higgs 
width can in principle be measured {\it directly} 
by tuning the collider to scan across the Higgs resonance.
One can also use the polarization of
the photon beams to measure various asymmetries in Higgs production
and decay, which are sensitive to the CP quantum number of the Higgs 
boson~\cite{asner}.

Finally, we note that substantial improvements are possible 
for precision measurements of 
$m_W$, $m_t$ and electroweak mixing angle measurements 
at the LC~\cite{futureprecision}.  
But, the most significant improvements can be achieved at
the GigaZ~\cite{gigaz}, 
where the linear collider operates at $\sqrt{s}=m_Z$ and
$\sqrt{s}\simeq 2m_W$.  With an
integrated luminosity of 50~fb$^{-1}$, one can collect
$1.5\times 10^9$ $Z$ events and about $10^6$ $W^+W^-$ pairs in the
threshold region.  Employing a global fit to the precision 
electroweak data in the
Standard Model, the anticipated fractional Higgs mass uncertainty
achievable would be about $8\%$.
This would provide a stringent test for the theory of the Higgs boson,
as well as very strong
constraints on any new physics beyond the Standard Model that couples
to the $W$ and $Z$ gauge bosons.

\subsection{Higgs Bosons in Supersymmetric Extensions of the Standard Model}

\begin{figure}[b!]
\begin{center}
\resizebox{\textwidth}{!}{
\unitlength1cm
\begin{picture}(16,7.1)
\put(0,-0.4){\includegraphics[height=3.223in,width=3.117in]{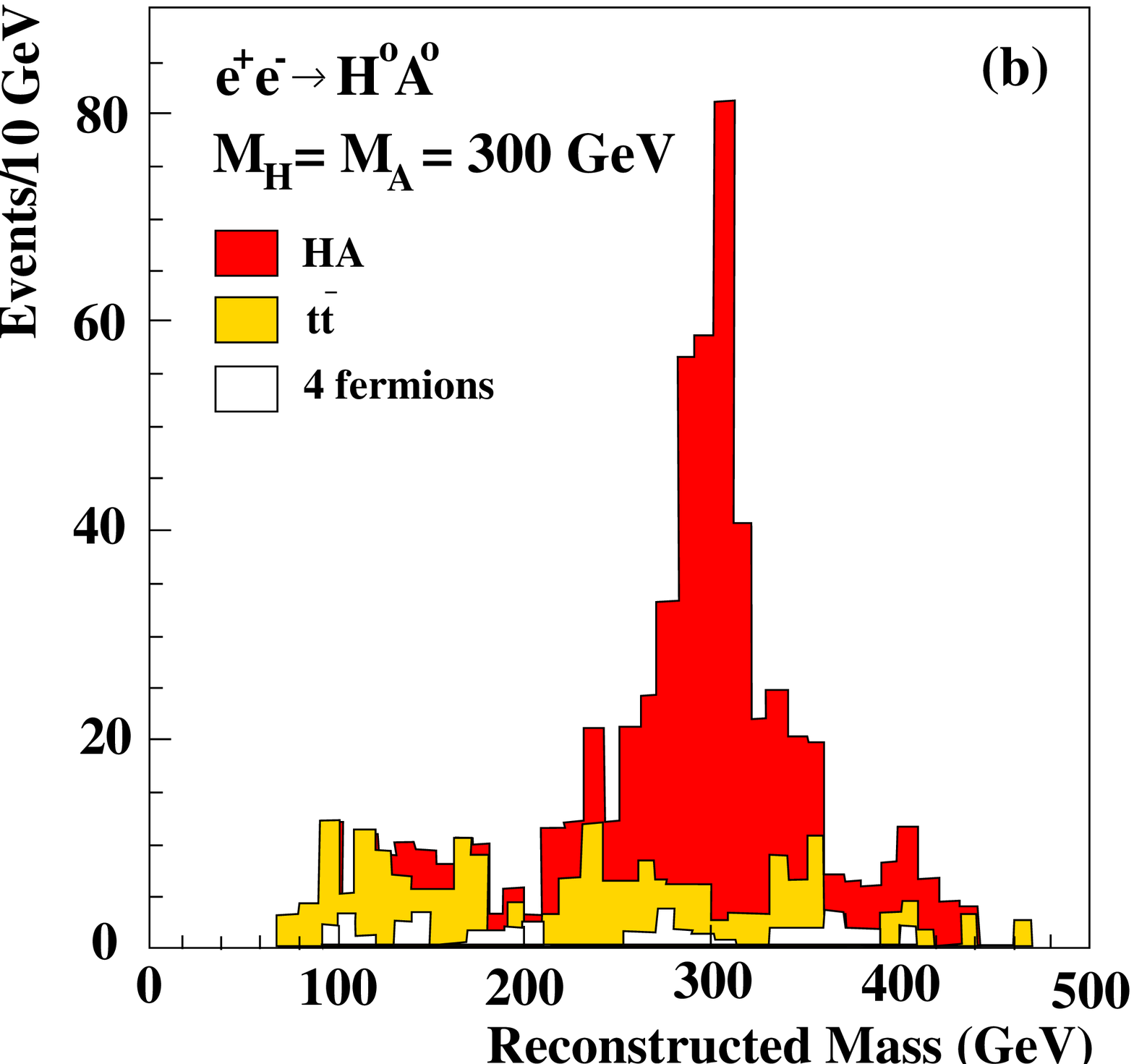}}
\put(8.2,-0.48){\includegraphics[height=3in,width=3in]{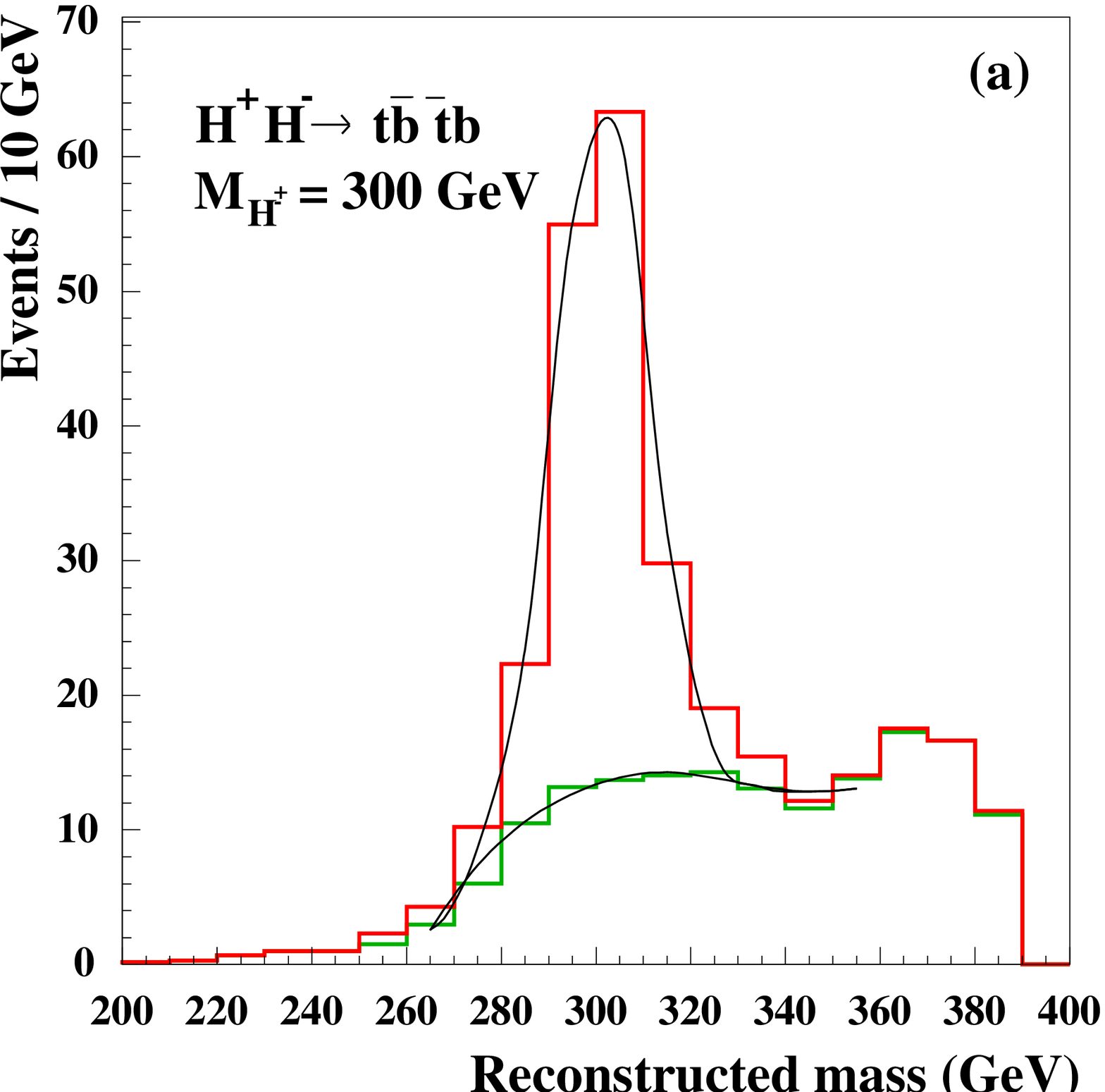}}
\end{picture}
}\end{center}
\caption{Heavy MSSM Higgs states at the LC for
$\sqrt{s}=800$~GeV~\cite{Tesla-TDR}: 
(a)~Reconstructed $H$ and $A$ mass peak from $e^+e^-\to HA\to b\bar b
b\bar b$ for 50 fb$^{-1}$ of data; and (b)~the dijet invariant mass
distribution for $e^+e^-\to H^+H^-\to t\bar b\bar t b$ candidates
after applying the intermediate $t$ and $W$ mass and the equal final
state mass constraints for 500 fb$^{-1}$ of data.}
        \label{fig:mssmhiggsproduction}
\end{figure}

We first focus on the 
production of $h$, $H$, $A$ and $H^\pm$ of the MSSM.
The main production mechanisms are (i)~single Higgs production 
($e^+e^-\to Zh$, $ZH$) via Higgs-strahlung, (ii)~associated
neutral Higgs pair production ($e^+e^-\to hA$, $HA$)
via $s$-channel $Z$ exchange, and (iii) charged Higgs pair production
($e^+e^-\to H^+H^-$).  Processes (i) and (ii) are complementary to
each other as a consequence of unitarity sum rules for
tree-level Higgs couplings~\cite{wudka}.  In particular, 
$g^2_{\phi ZZ} + 4m^2_Z g^2_{\phi AZ}=g^2 m_Z^2/\cos^2\theta_W$ (for
$\phi=h$, $H$), which shows that both $g^2_{\phi ZZ}$ and $g^2_{\phi
AZ}$ cannot simultaneously vanish.
For $m_A\gsim 200$~GeV, one finds that $m_A\sim m_H\sim m_{H^\pm}\gg m_h$
and $g_{HZZ}\sim g_{hAZ}\sim 0$, as a
consequence of the decoupling limit in which the properties of $h$
are nearly indistinguishable from those of the SM Higgs 
boson~\cite{decouplinglimit}.  
Thus, at the LC with center-of-mass energy $\sqrt{s}$, the
Higgs-strahlung of the lightest Higgs boson $Zh$ and pair production
of the heavy Higgs bosons $HA$ and $H^+H^-$ are dominant if
$m_A\lsim\sqrt{s}/2$.  In this case, the heavy Higgs 
states can be cleanly reconstructed at the
linear collider,  as seen in Figure~\ref{fig:mssmhiggsproduction}a 
and~\ref{fig:mssmhiggsproduction}b.
On the other hand, since $m_h\lsim 135$~GeV, 
a center-of-mass energy of 300 GeV is more than
sufficient to cover the entire MSSM parameter space with
certainty.  Thus, the
light Higgs boson, $h$, is accessible at the LC
while the observation of $H$, $A$ and
$H^\pm$ is possible only if $\sqrt{s}$ is sufficiently
large.
The heavier Higgs states could lie beyond the discovery reach
of the LC ($\sqrt{s} \leq 1$~TeV), and require a multi-TeV linear collider
for discovery and study.
In this case, the precision measurements of the light Higgs decay 
branching ratios and
couplings achievable at an $e^+e^-$ linear collider are critical for
distinguishing between $\hsm$ and $h$
of a non-minimal Higgs sector with properties close to that of the 
SM Higgs boson.

\begin{figure}[t!]
\begin{center}
\resizebox{\textwidth}{!}{
\includegraphics*[19,142][529,682]{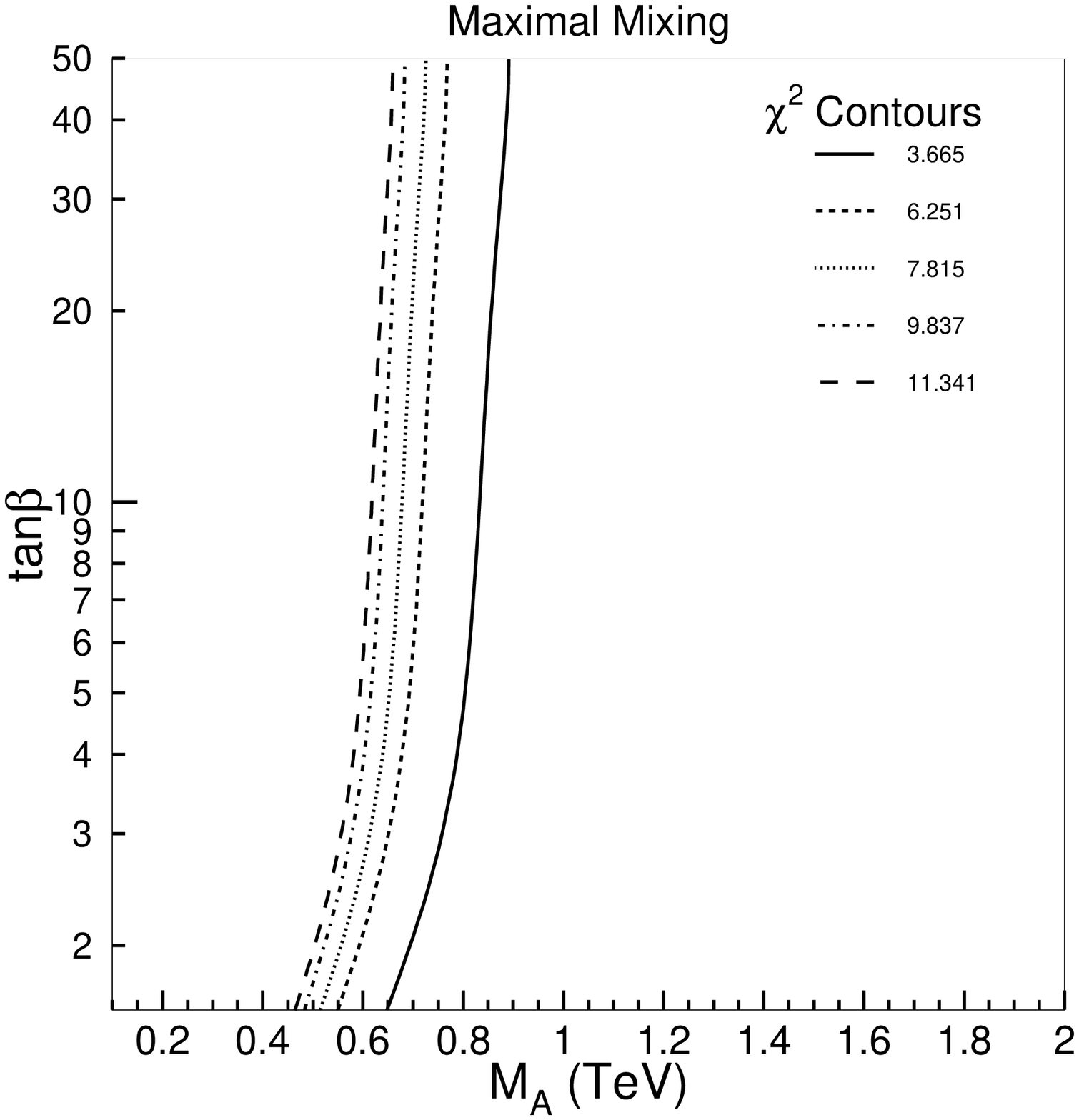}
\includegraphics*[19,142][529,682]{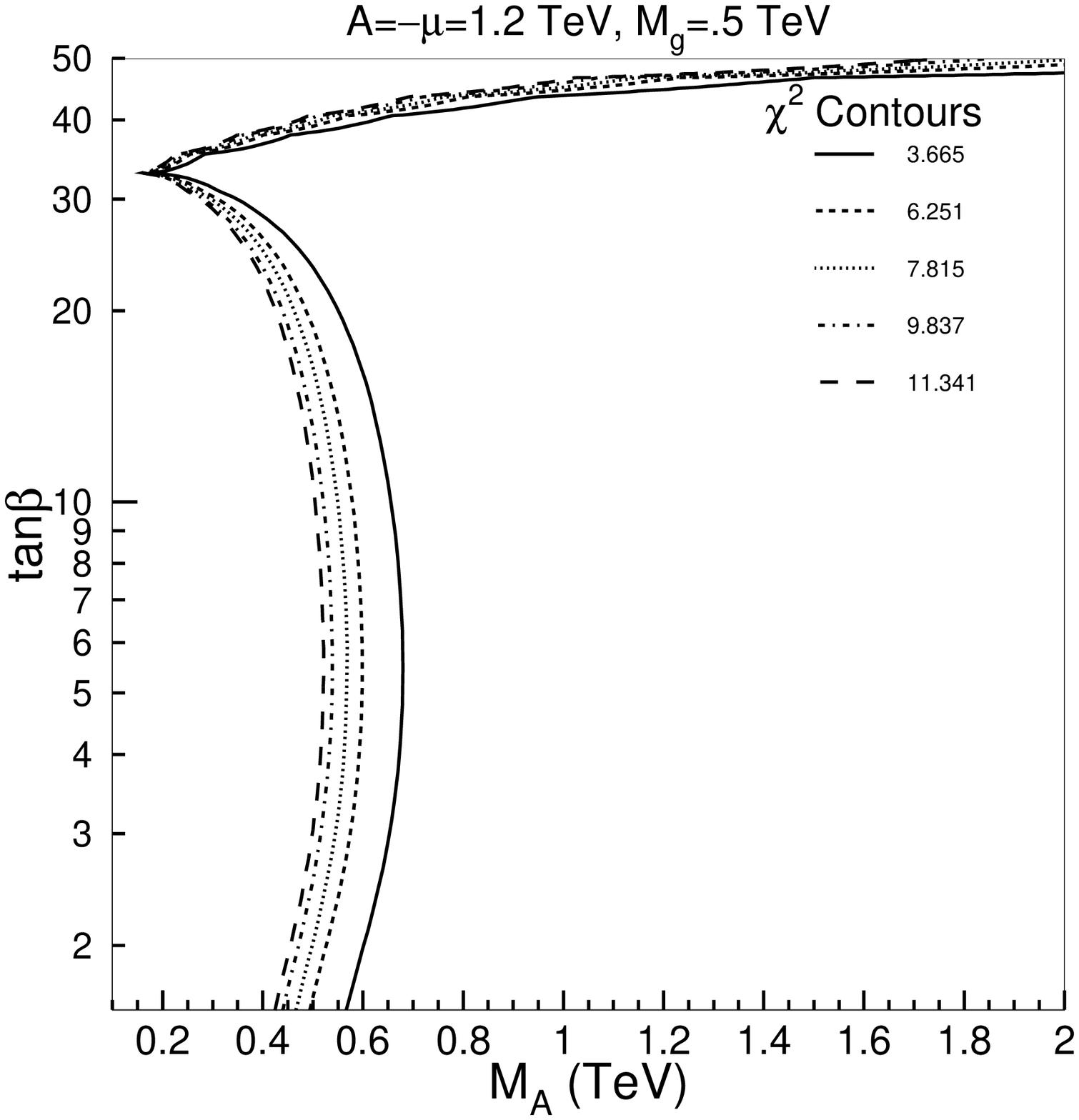}
}

\end{center}
\caption{Contours of $\chi^2$ for Higgs
boson decay observables for the maximal mixing scenario [left panel] and
for a different choice of MSSM parameters for which the
one-loop shift to the $hb\bar b$ coupling is large [right panel].  See
Ref.~\cite{chlm} for additional details.
The contours correspond to
68, 90, 95, 98 and 99\% confidence levels (right to left) for the three
observables $g^2_{hbb}$, $g^2_{h\tau\tau}$, and $g^2_{hgg}$.  }
\label{fig:chisquare}
\end{figure}

To illustrate the challenge of probing the decoupling limit, suppose
that $m_A>\sqrt{s}/2$ so that only the light Higgs boson, $h$, can be
observed directly at the LC.  However, in this region of
decoupling the deviation of the couplings of $h$ from those
of the SM Higgs boson approach zero.  In particular, the 
fractional deviation
scales as $m_Z^2/m_A^2$, so that if precision measurements reveal a
non-zero deviation, one could in principle derive a constraint on the
heavy Higgs masses of the model.  In the MSSM, the constraint can be
sensitive to the MSSM parameters which control the radiative
corrections to the Higgs couplings.  This is illustrated in
Figure~\ref{fig:chisquare}, where the constraints on $m_A$ are derived for
two different sets of MSSM parameter choices~\cite{chlm}.  
Here, a simulation of a
global fit of measured $hbb$, $h\tau\tau$ and $hgg$ couplings is made
and $\chi^2$ contours are plotted indicating the constraints in the
$m_A$--$\tan\beta$ plane assuming a deviation from SM Higgs couplings
is seen.  In the maximal mixing scenario, the constraints on $m_A$ are
significant and rather insensitive to the value of $\tan\beta$.
However in some cases, as shown in Figure~\ref{fig:chisquare}b,
a region of $\tan\beta$ may yield almost no constraint on $m_A$.
Of course, if supersymmetric particles are discovered 
prior to the precision Higgs measurements, additional information
about the MSSM spectrum can be employed to further refine the
analysis.

The $e^+e^-$ collider running in the $\gamma\gamma$ collider mode
presents additional opportunities for the study of the MSSM Higgs sector.
Resonance production $\gamma\gamma \to H$ and $A$
can be used to extend the reach in
Higgs masses beyond the limit set by $HA$ pair production in the
$e^+e^-$ mode~\cite{n41a,asner,velasco}.  Typically, one can probe the heavy
Higgs masses out to $m_A\sim 0.8\sqrt{s}$ (where $\sqrt{s}$ is the
center of mass energy of the LC).  This extends the MSSM Higgs search
to regions of the $m_A$--$\tan\beta$ parameter space for which the
LHC is not sensitive in general
(the so-called ``blind wedge'' of large $m_A$ and
moderate values of $\tan\beta$).  

As noted above, at 
least one Higgs boson must be observable at the LC in the MSSM.
In non-minimal supersymmetric models, additional Higgs bosons appear in
the spectrum, and the ``no-lose'' theorem of the MSSM must be
reconsidered.  For example, in the non-minimal supersymmetric
extension of the Standard Model (the so-called NMSSM where a Higgs
singlet is added to the model~\cite{nmssm}),
the lightest Higgs boson decouples
from the $Z$ boson if its wave function is dominated by the Higgs
singlet component.  However, in this case the second
lightest Higgs boson usually plays the role of $h$ of the MSSM.  That is,
the mass of the second lightest neutral CP-even
Higgs boson is light (typically below 150 GeV) with
significant couplings to the $Z$, so that it can be produced by the
Higgs-strahlung process with an observable cross section~\cite{okada}.
If the second lightest Higgs boson also decouples from the $Z$, then
the third lightest will play the role of the lightest CP-even Higgs
boson of the MSSM for which the observation is ensured, and so on.
Even in bizarre scenarios where all the neutral Higgs boson share equally
in the coupling to $ZZ$ (with the sum of all squared couplings 
constrained to equal the square of the $\hsm ZZ$ coupling~\cite{wudka}), 
the ``no-lose'' theorem still applies---Higgs production at the LC must be 
observable~\cite{jack}.  In contrast, despite significant progress,
there is no complete guarantee
that at least one Higgs boson of the NMSSM must be discovered 
at the LHC for all choices of the model parameters~\cite{nolose-lhc}.

One of the key parameters of the MSSM Higgs sector
is the value of the ratio of Higgs vacuum expectation values,
$\tan\beta$.  In addition to providing information about the structure
of the non-minimal Higgs sector, the measurement of this parameter
also provides an important check of supersymmetric structure, since
this parameter also enters the chargino, neutralino 
and third generation squark mass matrices and couplings.
Thus, $\tan\beta$ can be measured independently using 
supersymmetric processes and compared to the value obtained from
studying the Higgs sector.  Near the decoupling limit, the properties of
$h$ are almost indistinguishable from those of $\hsm$, and thus
no information can be extracted on the value of $\tan\beta$.
However, the properties of
the heavier Higgs bosons are $\tan\beta$-dependent.  
Far from the
decoupling limit, all Higgs bosons of the MSSM will be observable at
the LC and exhibit strong $\tan\beta$-dependence in their couplings.
Thus, to extract
a value of $\tan\beta$ from Higgs processes, one must observe the
effects of the heavier Higgs bosons of the MSSM at the LC.  

The ultimate accuracy of the $\tan\beta$ measurement at the LC depends
on the value of $\tan\beta$.  In Ref.~\cite{gunionetal}, it is argued that
one must use a number of processes, including $b\bar b b\bar b$ final
states arising from $b\bar b H$, $b\bar b A$, and $HA$ production,
and $t\bar t b\bar b$ final states arising from $t\bar b H^+$,
$b\bar t H^-$ and $H^+ H^-$ production.  One subtlety that arises here
is that in certain processes, the determination of $\tan\beta$ 
may be sensitive to loop corrections that depend on the values of
other supersymmetric parameters.  One must settle on a consistent 
definition of $\tan\beta$ when loop corrections are included
[analogous to the ambiguity in the definition of the one-loop electroweak
mixing angle].  A comprehensive analysis of the extraction of $\tan\beta$
from collider data, which incorporates loop effects, has not yet been given.

The study of the properties of the heavier MSSM Higgs bosons 
(mass, width, branching ratios, quantum numbers, {\it etc.}) 
provides a number of additional challenges.
For example, in the absence of CP-violation, 
the heavy CP-even and CP-odd Higgs bosons, $H$ and $A$, are
expected to be nearly mass-degenerate.  Their CP quantum numbers and
their separation can be investigated at the same time in the
photon-photon collider mode of the LC.  If linearly
polarized photons are used in parallel polarization states, only the
CP-even Higgs boson $H$ will be produced, while in perpendicular
polarization states only the CP-odd Higgs boson $A$ will be
produced. Thus, the CP quantum numbers and the separation of the two
different states can be achieved.
So far, we have implicitly assumed that the neutral Higgs bosons are
CP eigenstates.  In the MSSM, the Higgs-sector is CP-conserving at
tree-level.  But, in supersymmetric models with explicit CP violation, 
radiative corrections can induce nontrivial CP-mixing among the
neutral Higgs states~\cite{cpcarlos}.  In the 
decoupling limit, the lightest
Higgs boson, $h$, remains CP-even, while the two heavier Higgs states
mix and exhibit CP-violating interactions with fermions~\cite{ghk}.
In non-minimal supersymmetric extensions of the Standard Model,
the more complicated Higgs sectors can also exhibit CP-violating
properties.  In the case of a CP-violating Higgs sector,
the observation and measurement
of the Higgs bosons become much more challenging, and an
$e^+e^-$ collider can uniquely test the nature of the couplings of the Higgs
neutral eigenstates of mixed CP parity to gauge bosons and fermions.

\subsection{Strong Electroweak Symmetry Breaking Dynamics}

\begin{figure}[b!]
\vspace{-6mm}
\begin{center}
\resizebox{0.85\textwidth}{!}{
\includegraphics*[scale=0.02,angle=-90]{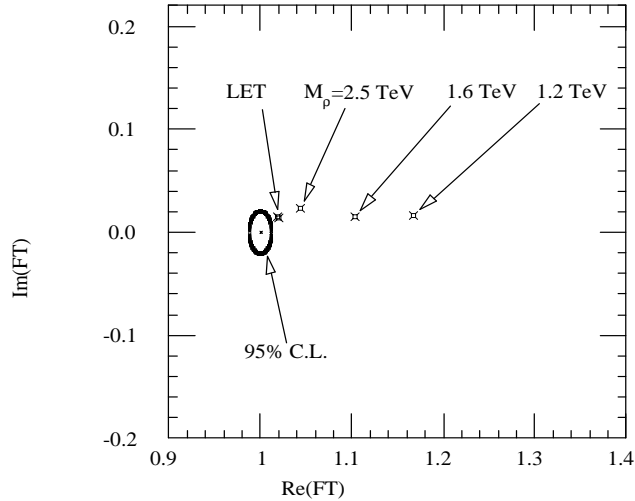}
}
\end{center}
\vspace{-2mm}
\caption{The $W_L W_L$ scattering form factor $F_T$, 
for various masses $M_{\rho}$ of a new
vector resonance in $e^+e^- \to W^+W^-$.
The strong threshold effects in $WW$ scattering, 
based on taking the $WW$ amplitude to be exactly given by the
amplitude as predicted by the low-energy theorem (LET), is indicated
by the LET point.
The contour is indicative for the precision attainable 
with 500~fb$^{-1}$ at $\sqrt{s}=500$~GeV.  Taken from~\cite{bark}.
}
\label{fig:form-factor-strongewsb}
\end{figure}

Important steps in exploring strong electroweak symmetry breaking can
be taken already at the LC with $\sqrt{s}\geq 500$~GeV and an
integrated luminosity of 500 fb$^{-1}$ and above.  Even if the masses of 
new heavy resonances associated with the symmetry breaking sector 
are in the TeV range, their effects can be indirectly observed
at the LC with $\sqrt{s}\leq 1$~TeV.  In
$e^+e^-\to W^+ W^-$, the entire threshold region for
the onset of the new strong interactions can be covered up to
scales of 3 TeV~\cite{bark}.
Strong quasi-elastic $WW$ scattering, 
the $W$ bosons emitted from the
high-energy electron and positron beams, can be studied
directly up to scales of the same size~\cite{n43a}.
Isospin-zero resonance
channels as well as isospin-two exotic channels are accessible in this
way. New $\rho$-type resonances can be studied as virtual states for
masses up to several TeV, 
as illustrated in Figure~\ref{fig:form-factor-strongewsb}. 
Pseudo-Goldstone bosons may be
accessible in $e^+e^-$ annihilation and $\gamma\gamma$ collisions up to a
few hundred GeV~\cite{casalbuoni,Lane:2002wb}.

Strong gauge boson interactions also can induce anomalous triple and
quartic gauge couplings at tree-level.  Both CP-conserving and
CP-violating couplings are possible.  For example, precision
measurements of the process $e^+e^-\to W^+W^-$ are sensitive to anomalous
contributions to the static magnetic and electric dipole and
quadrupole moments.  The expected errors in the anomalous couplings,
relative to the Standard Model triple gauge boson coupling, range
from $10^{-4}$ to $10^{-3}$ at the LC with $\sqrt{s}=500$~GeV to 1~TeV
and an integrated luminosity of $0.5~{\rm to }~1~{\rm ab}^{-1}$.  At
these accuracies, one can begin to probe the contributions to the
anomalous couplings from Standard Model (or MSSM) perturbative
one-loop corrections.  Corrections due to the strong electroweak
symmetry breaking sector are likely to be of the same order of
magnitude, or perhaps somewhat larger, and they can provide independent
evidence for the existence of new TeV-scale physics.

A multi-TeV $e^+e^-$ collider is an excellent tool to study new strong
interaction resonances in great detail. Since $W$ bosons can be
reconstructed in the jet decay channels, 
the dynamics of the new resonances can be explored in a 
more comprehensive way than at
hadron colliders. Such a machine is the appropriate instrument for
fully developing the picture of the
new strong forces in the electroweak sector.

\section{EWSB Physics at Far-Future Collider Facilities}

\subsection{Probing EWSB at a $\mu^+\mu^-$ Collider}

\begin{figure}[b!]
\begin{center}
\unitlength1cm
\begin{picture}(16,8.0)
\put(-0.3,-0.2){\includegraphics*[width=3.193in,height=3.3in]{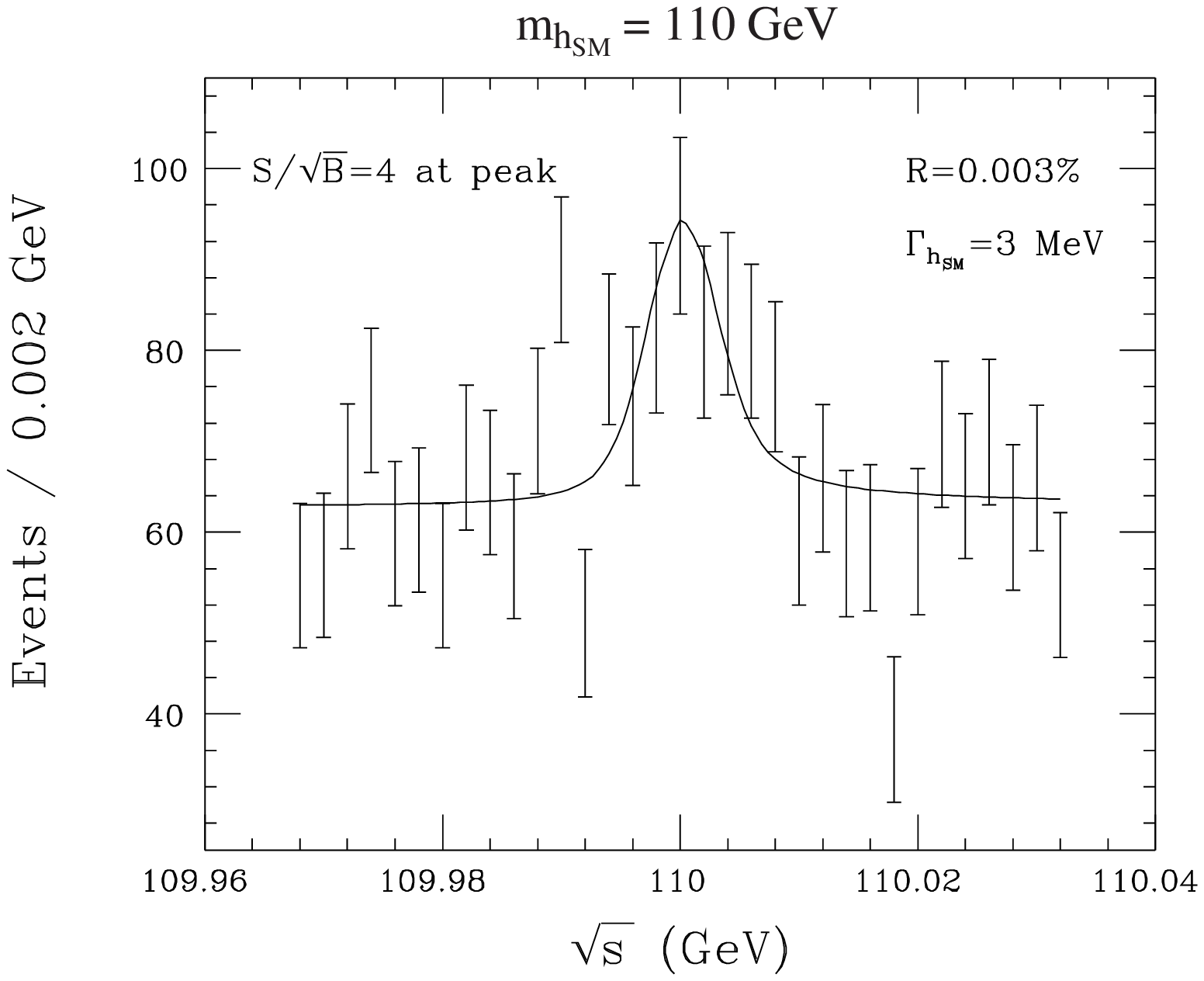}}
\put(7.9,-0.33){\includegraphics*[width=3.3in,height=3.396in]{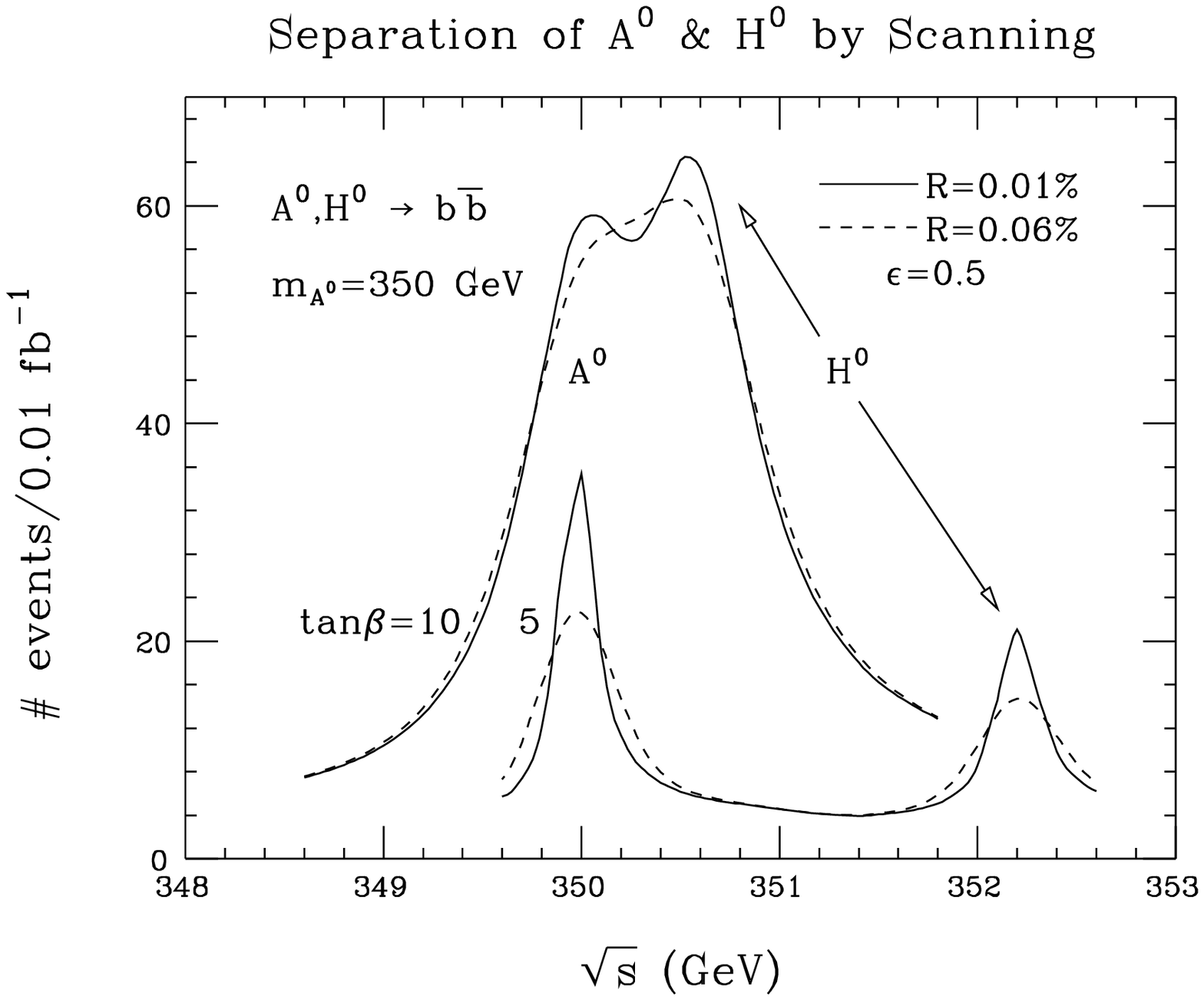}}
\end{picture}\\[-2mm]
\end{center}
        \caption{Higgs-boson signals at a muon collider,
     taken from Refs.~\cite{ankenbrandt} and \cite{Barger:1996jm}.  
     (a)~Scan of the 
     $s$-channel Higgs resonance in the Standard Model
    for $\mhsm = 110$~GeV assuming a beam energy width of 
    $R=0.003\%$ and 1.5~pb$^{-1}$
    per scan point. (b)~Resolution of $H$--$A$ splitting
    in supersymmetric theories.}
        \label{fig:mupmumtohiggs}
\end{figure}

In contrast to Higgs production at
electron-positron colliders, the Higgs boson can be produced as
an $s$-channel resonance in a $\mu^+\mu^-$ 
collider~\cite{Barger:1996jm,ankenbrandt,mumu-cern,bbgh,n55a,mufactory} 
with an appreciable rate, since the Higgs coupling to
muons is sufficiently large to generate a sizeable production cross
section. For a Higgs boson mass in the lower part of the intermediate
mass range, roughly $10^4$ particles can be produced in a few years, with the
same number of background events in the $b\bar b$ channel.
Given the expected energy resolution, the Higgs mass can be
measured in such a machine with the accuracy of a few MeV, as shown in
Figure~\ref{fig:mupmumtohiggs}a, similar to the precision of the $Z$ mass 
measurement at LEP.  The Higgs width can be measured directly 
from a scan of the Higgs lineshape, with an accuracy of order $20\%$. 
Since the Higgs-boson width becomes rapidly wider at the upper end of
the intermediate mass range, the
Higgs resonance-signal is no longer observable at the $\mu^+\mu^-$ collider 
for $\mhsm\gsim 160$~to~180~GeV.
Anticipating the
discovery of a fundamental relation between the Higgs mass and the
$Z$ mass in a future comprehensive theory of particle physics, the
high precision with which the Higgs boson mass can be measured at a
muon collider could turn out to be a critical aspect in testing such
a theory, in analogy to the relation between the $Z, W^{\pm}$ masses
and the electroweak mixing angle in the Standard Model.

The sharp energy of a muon collider can be exploited to
resolve the nearly mass-degenerate CP-even Higgs boson $H$ and the
CP-odd Higgs boson $A$ in supersymmetric theories, as shown in 
Figure~\ref{fig:mupmumtohiggs}b.
Clearly, many other aspects of the Higgs sector can be
studied at a $\mu^+\mu^-$ collider~\cite{Barger:1996jm,mufactory}. 
In particular,
if polarized beams are available, one could explore the CP quantum numbers 
of the Higgs boson(s) or probe CP-violation in the Higgs sector.
There are several CP-violating observables, which are unique to 
$s$-channel Higgs production at the $\mu^+\mu^-$ collider, that 
can be constructed using muon polarization 
vectors~\cite{Atwood:1995uc,Barger:1996jm}
and/or three-momenta and spins of the final 
particles~\cite{Kramer:1993jn}.
These asymmetries are degraded for partially polarized muon
beams and by the effects of the precession of the spins of the colliding
beams.  Nevertheless, in some cases,
the CP quantum number of the SM Higgs boson 
or of the neutral Higgs bosons of an extended Higgs sector can be extracted
with reasonable accuracy~\cite{pliszka,mufactory}
({\it e.g.}, for the MSSM Higgs sector with large
radiatively-induced CP-violating Higgs couplings~\cite{cpcarlos}).

Of course, the Higgs boson is also produced via the same Higgs-strahlung and
vector-boson fusion processes that operate at $e^+e^-$ colliders.
Thus, much of the LC program for Higgs physics is also possible at a
$\mu^+\mu^-$ collider.
However, (presumably) reduced luminosities at the $\mu^+\mu^-$ collider and
backgrounds due to the decaying muons will degrade some of the LC precision
Higgs measurements previously discussed.

\subsection{Probing EWSB at a Very Large Hadron Collider (VLHC)}

If strong electroweak symmetry breaking, characterized by a scale of several
TeV, is realized in nature, a proton collider with
energies far above that of the LHC~\cite{VLHC} will be a crucial instrument,
complementary to multi-TeV lepton colliders,
to study the dynamics of the system.
The significance of quasi-elastic $WW$ scattering signaling either the
onset of the new strong interactions or the formation of new
resonances, is greatly enhanced compared to the LHC, and it provides a
motivation for detailed experimental studies.

\begin{figure}[b!]
\begin{center}
\includegraphics[height=3.2in]{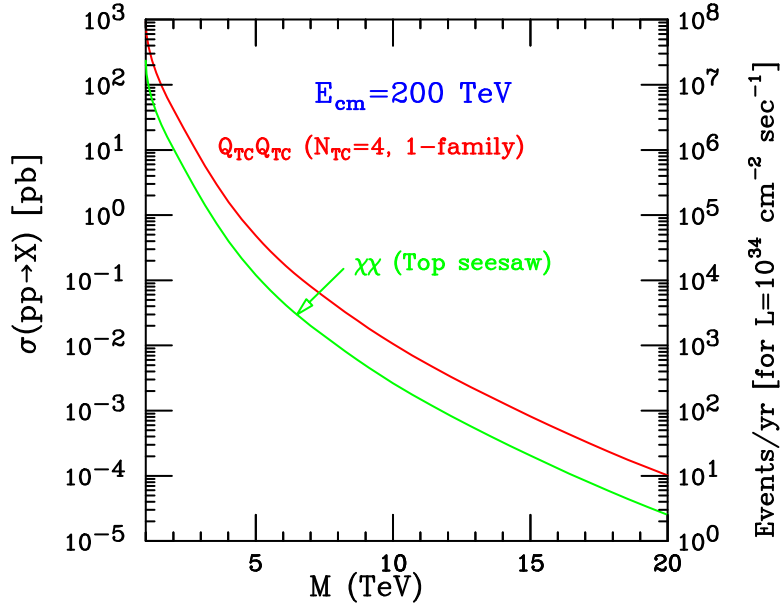}
\end{center}
  \caption{The cross section for technicolor $Q_{\rm TC}$
   pair production
   (solid line) and pair production of SU(2)$_L$ singlet top-quark
   partners, $\chi\ls{L}$ and $\chi\ls{R}$, in top-color models (dashed
   line) at the VLHC.  The calculation assumes one degenerate
   isodoublet of techniquarks, and $\chi\ls{L}$ and
   $\chi\ls{R}$ are taken degenerate in mass.
   The right vertical scale
   shows the number of events per year, assuming a total yearly
   integrated luminosity of 100~fb$^{-1}$. Taken from Ref.~\cite{baur}.} 
   \label{fig:technistuff}
\end{figure}

If the Higgs boson is a composite particle, a proton collider with very
high energies may be a unique instrument 
to search for its
constituents. Examples include technicolor and top-color theories
in which the new quarks may have masses of several TeV and above.
For example, a condensate of $\overline Q_{\rm TC} Q_{\rm TC}$
may be responsible for electroweak symmetry breaking, where
$Q_{\rm TC}$ are
techniquarks which make up the fundamental representation of an SU(4)
technicolor group~\cite{technireview}.
In the top-seesaw model of Ref.~\cite{topseesaw}, the top quark and a novel
weak SU(2)$_L$ singlet quark $\chi$
are responsible for the dynamical breaking of electroweak
symmetry.  The cross sections for production of these new particles at
the VLHC are shown in Figure~\ref{fig:technistuff}~\cite{baur}.
However, detailed experimental studies of the signals and backgrounds 
in the hadronic environment are needed before firm conclusions can be drawn.

\section{Conclusions}

The physical origin of electroweak symmetry breaking is not yet known.
In all theoretical approaches and models, 
the dynamics of electroweak symmetry
breaking must be revealed at the TeV-scale or below.  This energy
scale will be thoroughly explored by hadron colliders, starting with
the Tevatron and followed later in this decade by the LHC.
Even though the various theoretical alternatives can only be confirmed
or ruled out by future collider experiments,
a straightforward interpretation of the electroweak precision data
suggests that electroweak symmetry breaking dynamics is
weakly-coupled, and a Higgs boson with mass between
100 and 200 GeV must exist.   With the
supersymmetric extension of the Standard Model, this interpretation
opens the route to grand unification of all the fundamental forces, with the
eventual incorporation of gravity in particle physics.
The observation of a light Higgs boson at the Tevatron or the LHC is
the crucial first step. However, a high-luminosity $e^+e^-$ collider,
now under development, is
needed to clarify the nature of the Higgs boson in a comprehensive form
and to establish scalar sector dynamics as the mechanism 
{\it sui generis} for generating
the masses of the fundamental particles.  If 
strong electroweak symmetry
breaking dynamics is realized in nature, supporting evidence can
initially be extracted from experiments at the LHC and at an
$e^+e^-$ linear collider with $\sqrt{s}=500$~GeV---1~TeV, but the new strong
interaction sector can only be fully explored at multi-TeV
lepton and proton facilities.

In summary, discovering and interpreting new phenomena require energy
frontier facilities and high precision capabilities.  The search for
the dynamics of electroweak symmetry breaking calls for colliders that
probe energy scales from a few hundred GeV up to a TeV.  Theoretical
explanations of the mechanism of electroweak symmetry breaking demand
new physics beyond the Standard Model at or near the TeV scale.  There
are fundamental questions concerning electroweak symmetry breaking and
physics beyond the Standard Model that cannot be answered without a
high energy physics program at an $e^+e^-$ linear collider overlapping
that of the LHC.  Discoveries at these machines will elucidate the TeV
scale, and they will pave the way for facilities that will explore new
and higher energy frontiers at the multi-TeV scale.

\begin{acknowledgments}

We would like to thank Jack Gunion, Sven Heinemeyer and Tom Rizzo for
their careful reading of the manuscript and a number of useful
suggestions.  We also acknowledge fruitful
conversations with JoAnne Hewett, Carlos Wagner, Georg Weiglein and 
Dieter Zeppenfeld.  In addition,
we greatly appreciate the contributions of the members of the Electroweak
Symmetry Breaking (P1) working group participants, with special thanks to
the P1 plenary speakers and subgroup conveners for their efforts.
Finally, we are particularly grateful to Chris
Quigg and the Snowmass Workshop organizing committee for inviting us to
convene the P1 working group and for providing such a stimulating atmosphere.  

Fermilab is operated by Universities Research Association, Inc.\
under contract no.~DE-AC02-76CH03000 with the U.S. Department of
Energy.  D.W.G. is supported in part by an NSF Career award
no.~PHY-9818097 and in part by the U.S. Department of Energy
under grant no.~DE-FG02-95ER20899.
H.E.H. is supported in part by the U.S. Department of Energy
under grant no.~DE-FG03-92ER40689.  A.S.T. acknowledges the support of
U.S. Department of Energy contract no.~DE-AC02-98CH10886.

\end{acknowledgments}

\end{document}